\documentclass{aa}
\usepackage[varg]{txfonts}
\usepackage{hyperref}

\usepackage{natbib}
\bibpunct{(}{)}{;}{a}{}{,} 

\usepackage{graphicx}

\begin{document}

\title{Detectability of atmospheric features of Earth-like planets in the habitable zone around M dwarfs}

\author{Fabian Wunderlich\inst{1}
   \and Mareike Godolt\inst{1}
   \and John Lee Grenfell\inst{2}
   \and Steffen Städt \inst{3}
   \and Alexis M. S. Smith\inst{2}
   \and Stefanie Gebauer \inst{2}
   \and Franz Schreier \inst{3}
   \and Pascal Hedelt \inst{3}
   \and Heike Rauer\inst{1,2,4}}

\institute{Zentrum für Astronomie und Astrophysik, Technische Universität Berlin, Hardenbergstraße 36, 10623 Berlin, Germany\\ \email{fabian.wunderlich@tu-berlin.de}
  \and Institut für Planetenforschung, Deutsches Zentrum für Luft- und Raumfahrt, Rutherfordstraße 2, 12489 Berlin, Germany 
  \and Institut für Methodik der Fernerkundung, Deutsches Zentrum für Luft- und Raumfahrt,  82234 Oberpfaffenhofen, Germany
  \and Institut für Geologische Wissenschaften, Freie Universität Berlin, Malteserstr. 74-100, 12249 Berlin, Germany}

\date{}

\abstract {The characterisation of the atmosphere of exoplanets is one of the main goals of exoplanet science in the coming decades.}
{We investigate the detectability of atmospheric spectral features of Earth-like planets in the habitable zone (HZ) around M dwarfs with the future James Webb Space Telescope (JWST).} 
{We used a coupled 1D climate-chemistry-model to simulate the influence of a range of observed and modelled M-dwarf spectra on Earth-like planets. The simulated atmospheres served as input for the calculation of the transmission spectra of the hypothetical planets, using a line-by-line spectral radiative transfer model. To investigate the spectroscopic detectability of absorption bands with JWST we further developed a signal-to-noise ratio (S/N) model and applied it to our transmission spectra.} 
{High abundances of methane (CH$_4$) and water (H$_2$O) in the atmosphere of Earth-like planets around mid to late M dwarfs increase the detectability of the corresponding spectral features compared to early M-dwarf planets. Increased temperatures in the middle atmosphere of mid- to late-type M-dwarf planets expand the atmosphere and further increase the detectability of absorption bands. To detect CH$_4$, H$_2$O, and carbon dioxide (CO$_2$) in the atmosphere of an Earth-like planet around a mid to late M dwarf observing only one transit with JWST could be enough up to a distance of 4~pc and less than ten transits up to a distance of 10~pc. As a consequence of saturation limits of JWST and less pronounced absorption bands, the detection of spectral features of hypothetical Earth-like planets around most early M dwarfs would require more than ten transits. We identify 276 existing M dwarfs (including GJ~1132, TRAPPIST-1, GJ~1214, and LHS~1140) around which atmospheric absorption features of hypothetical Earth-like planets could be detected by co-adding just a few transits.} 
{The TESS satellite will likely find new transiting terrestrial planets within 15~pc from the Earth. We show that using transmission spectroscopy, JWST could provide enough precision to be able to partly characterise the atmosphere of TESS findings with an Earth-like composition around mid to late M dwarfs.}

\maketitle 

\section{Introduction}

Dozens of terrestrial planets in the habitable zone (HZ) have been found so far. Most of these planets orbit cool host stars, the so-called M dwarfs\footnote{phl.upr.edu/projects/habitable-exoplanets-catalog}. The Transiting Exoplanet Survey Satellite (TESS) is expected to find many more of these systems in our solar neighbourhood in the near future \citep{ricker2014,sullivan2015,barclay2018}. Whether these planets can have surface conditions to sustain (complex) life is still debated \citep[e.g.][]{tarter2007,shields2016}. Because of the close-in HZ, M-dwarf planets are likely tidally locked \citep[e.g.][]{kasting1993,selsis2008}. To allow for habitable surface conditions they require a mechanism to redistribute the heat from the dayside to the nightside \citep[e.g. atmospheric or ocean heat transport;][]{joshi1997,joshi2003,selsis2008,yang2013,hu2014}. 
\begin{table*}
\centering
\caption{Stellar parameters for the Sun and all M dwarfs used in this study. Stars labelled with an asterisk are active M dwarfs. Effective temperatures of the stars which are included in the MUSCLES database ($T_{\text{eff}}$[K] MUSC.) are taken from \citet{loyd2016}. }              
\label{table:stars}      
\centering                                     
\begin{tabular}{l l llllll}          
\hline\hline                        
Star            & Type          & $T_{\text{eff}}$[K] Lit.      & $T_{\text{eff}}$[K] MUSC.           & $R/R_\text{$\odot$}$  & $M/M_\text{$\odot$}$                  & $L/L_\text{$\odot$}$ [10$^{-3}$] & $d$ [pc]
 \\    
\hline                                  
Sun             & G2            & 5772                  & -                             & 1.000                   & 1.000                                 & 1000.00               & 0.00 \\
GJ 832          & M1.5$^a$      & 3657$^b$              & 3816$\pm$250$^c$                 & 0.480$^a$             & 0.450$\pm$0.050$^a$                   & 26.00$^d$               & 4.93$^a$ \\
GJ 176          & M2$^e$        & 3679$\pm$77$^e$       & 3416$\pm$100$^c$                 & 0.453$\pm$0.022$^e$   & 0.450$^e$                             & 33.70$\pm$1.80$^e$      & 9.27$^e$ \\
GJ 581          & M2.5$^f$      & 3498$\pm$56$^f$       & 3295$\pm$140$^c$                 & 0.299$\pm$0.010$^f$   & 0.300$^f$                             & 12.05$\pm$0.24$^f$      & 6.00$^f$ \\
GJ 436          & M3$^g$        & 3416$\pm$56$^g$       & 3281$\pm$110$^c$                 & 0.455$\pm$0.018$^g$   & 0.507$\substack{+0.071\\-0.062}$ $^g$ & 25.30$\pm$1.20$^g$       & 10.23$^h$ \\
GJ 644          & M3*$^i$       & 3350$^j$              & -                             & 0.678$^k$               & 0.416$\pm$0.006$^i$                   & 26.06$^j$             & 6.50$^k$ \\
AD Leo          & M3.5*$^b$     & 3380$^b$              & -                             & 0.390$^l$               & 0.420$^l$                             & 24.00$^m$             & 4.89$^b$ \\
GJ 667C         & M3.5$^n$      & 3350$^o$              & 3327$\pm$120$^c$              & 0.460$^p$               & 0.330$\pm$0.019$^o$                   & 13.70$^q$             & 6.80$^q$ \\
GJ 876          & M4$^e$        & 3129$\pm$19$^e$       & 3062$\substack{+120\\-130}$ $^c$& 0.376$\pm$0.006$^e$       & 0.370$^e$                             & 12.20$\pm$0.20$^e$      & 4.69$^e$ \\
GJ 1214         & M4.5$^r$      & 3252$\pm$20$^s$       & 2935$\pm$100$^c$                 & 0.211$\pm$0.011$^s$   & 0.176$\pm$0.009$^s$                   & 4.05$\pm$0.19$^s$       & 14.55$\pm$0.13$^s$ \\
Proxima Cen.    & M5.5$^t$      & 3054$\pm$79$^t$       & -                             & 0.141$\pm$0.007$^t$     & 0.118$^t$                             & 1.55$\pm$0.02$^t$     & 1.30$^r$ \\
TRAPPIST-1      & M8$^u$        & 2559$\pm$50$^u$       & -                             & 0.117$\pm$0.004$^u$     & 0.080$\pm$0.007$^u$                   & 0.52$\pm$0.03$^u$     & 12.10$\pm$0.40$^u$ \\

\hline                                             
\end{tabular}
\tablebib{
(a)~\citet{bailey2008}; (b) \citet{gautier2007}; (c) \citet{loyd2016}; (d) \citet{bonfils2013}; (e) \citet{vonbraun2014};
(f) \citet{vonbraun2011}; (g) \citet{vonbraun2012}; (h) \citet{butler2004}; (i) \citet{segransan2000}; (j) \citet{reid1984}; 
(k) \citet{giampapa1996}; (l) \citet{reiners2009}; (m) \citet{pettersen1981}; (n) \citet{neves2014}; (o) \citet{anglada2013a}; 
(p) \citet{kraus2011}; (q) \citet{vanleeuwen2007}; (r) \citet{lurie2014}; (s) \citet{anglada2013b}; (t)~\citet{boyajian2012}; (u)~\citet{gillon2017}
}
\end{table*}
Another drawback of a planet lying close to its host star is the high luminosity during the pre-main-sequence phase \citep[e.g.][]{ramirez2014,luger2015,tian2015} or that it might be subject to strong stellar cosmic rays \citep[see e.g.][]{griessmeier2005,segura2010,tabataba2016,scheucher2018}. On the other hand M-dwarf planets are favourable targets for the characterisation of their atmosphere. The high contrast ratio and transit depth of an Earth-like planet around cool host stars favours a detection of spectral atmospheric features.
Hence, planets transiting M dwarfs are prime targets for future telescopes such as the James Webb Space Telescope \citep[JWST;][]{gardner2006,deming2009} and the European Extremely Large Telescope \citep[E-ELT;][]{gilmozzi2007,marconi2016}. \\
Additionally model simulations show that planets with an Earth-like composition can build up increased amounts of biosignature gases like ozone (O$_3$) and related compounds like water (H$_2$O) and methane (CH$_4$), which further increases their detectability \citep{segura2005,rauer2011,grenfell2013,rugheimer2015}. Knowing the ultraviolet radiation (UV) of the host star is crucial to understand the photochemical processes in the planetary atmosphere \citep[see e.g.][]{grenfell2014,rugheimer2015}. M dwarfs vary by several orders in the UV flux from inactive to active stars \citep{buccino2007,france2016}. \\
The approach of this study is to investigate the influence of a range of spectra of active and inactive stars with observed and modelled UV radiation on Earth-like planets in the HZ around M dwarfs. We furthermore aim to determine whether the resulting planetary spectral features could be detectable with JWST. 
Previous studies showed that multiple transits are needed to detect any spectral feature of an Earth-like planet with transmission or emission spectroscopy \citep{rauer2011,vonparis2011,barstow2016,barstow_irwin2016}. \citet{rauer2011} found that CH$_4$ of an Earth-like planet around AD~Leo (at 4.9~pc) would be detectable by co-adding at least three transits using transmission spectroscopy with JWST. To detect O$_3$ at least ten transits would be required. For the TRAPPIST-1 planets c and d at 12.1~pc, 30 transits are needed to detect an Earth-like concentration of O$_3$ with JWST \citep{barstow_irwin2016}. \\
In this study we first apply a 1D atmosphere model to calculate the global and annual mean temperature profiles and the corresponding chemical profiles of Earth-like planets orbiting M dwarfs. As model input we use 12 stellar spectra including the spectra of the MUSCLES database with observed UV radiation \citep{france2016}. In contrast to \citet{rugheimer2015} we use additional M-dwarf spectra and calculate transmission spectra of each Earth-like planet around M dwarfs with a radiative transfer model. We concentrate on transmission spectroscopy owing to the low chance of finding detectable absorption bands for temperate Earth-like planets with emission spectroscopy \citep[see][]{rauer2011,rugheimer2015, batalha2018}.
We further develop a model to calculate the signal-to-noise ratio (S/N) of planetary spectral features with any kind of telescope and apply this to the modelled spectra and up-to-date JWST specifications. With the S/N we evaluate the potential to characterise the atmosphere of Earth-like exoplanets around M dwarfs.  \\
The paper is organised as follows: In Section 2 we describe the atmospheric model, the scenarios, the line-by-line spectral model, and the S/N calculation. In Section 3 we first show the results of the atmospheric modelling and the resulting transmission spectra. Then we show the results of the S/N calculations. Section 4 presents the conclusions of this study.

\section{Methods and models}

\subsection{Climate-chemistry model}
We use a coupled 1D steady-state, cloud-free, radiative-convective photochemical model. The code is based on the climate model of \citet{kasting1984} and the photochemical model of \citet{pavlov2002} and was further developed by \citet{segura2003}, \citet{vonparis2008}, \citet{vonparis2010}, \citet{rauer2011}, \citet{vonparis2015}, \citet{gebauer2017}, \citet{gebauer2018b}, and \citet{gebauer2018}. The atmosphere in the climate module is divided into 52~pressure layers and the chemistry model into 64~equidistant altitude levels.\\
The radiative transfer of the climate module is separated into a short wavelength region from 237.6~nm to 4.545~$\mu$m with 38~wavelength bands for incoming stellar radiation and a long wavelength region from 1~$\mu$m to 500 $\mu$m in 25~bands for planetary and atmospheric thermal radiation \citep{vonparis2015}.
We consider Rayleigh scattering by N$_2$, O$_2$, H$_2$O, CO$_2$, CH$_4$, H$_2$, He, and CO using the two-stream radiative transfer method based on \citet{toon1989}. Molecular absorption in the short wavelength range is considered for the major absorbers H$_2$O, CO$_2$, CH$_4$, and O$_3$. Molecular absorption of thermal radiation by H$_2$O, CO$_2$, CH$_4$, and O$_3$ and continuum absorption by N$_2$, H$_2$O, and
CO$_2$ are included \citep{vonparis2015}. 
\\
Our chemistry module includes 55 species with 217 chemical reactions. An update compared to previous model versions, for example used in \citet{keles2018}, is to consider altitude dependent CO$_2$, O$_2$, and N$_2$ profiles instead of using an isoprofile as described in \citet{gebauer2017}. The photolysis rates are calculated within the wavelength range between 121.4 nm and 855~nm. For the effective O$_2$ cross sections in the Schumann-Runge bands we use the values from \citet{murtagh1988} as described in \citet{gebauer2018}. The mean solar zenith angle is set to 54$^\circ$ in the photochemistry module in order to best reproduce the 1976 U.S. Standard Atmosphere \citep{anderson1986}. The water vapour concentrations in the troposphere are calculated using the relative humidity profile of the Earth taken from \citet{manabe1967}.\\

\subsection{Stellar Input}
\begin{figure*}
\centering
   \includegraphics[width=17cm]{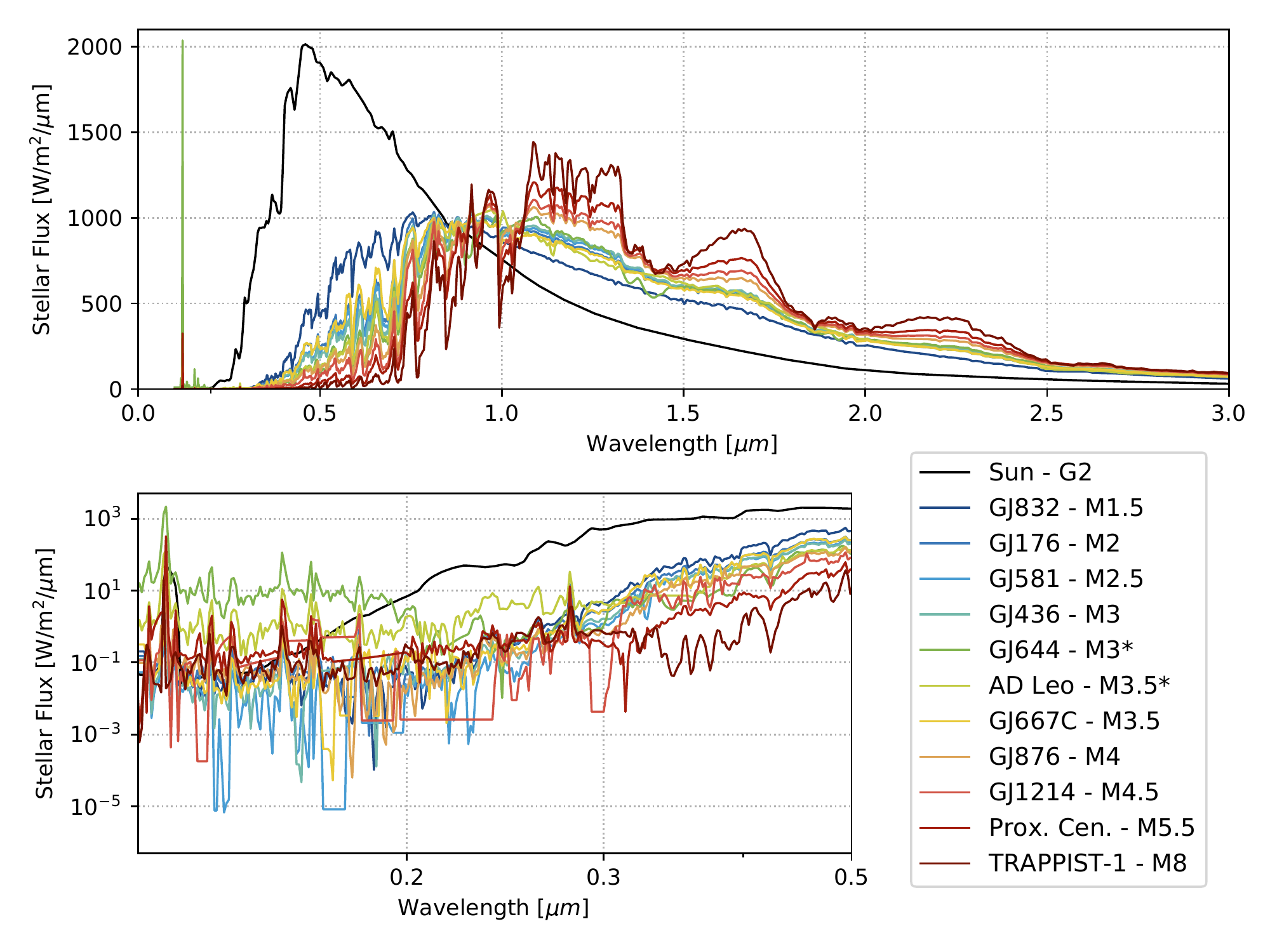}
     \caption{Stellar input spectra, scaled to reach a surface temperature of 288.15~K, of the modelled planet and binned to a resolving power of 200. AD Leo and GJ~644 are taken from the VPL website \citep{segura2003}, TRAPPIST-1 from \citet{omalley2017}, the other M-dwarf spectra from the MUSCLES database \citep{france2016}, and the solar spectrum from \citet{gueymard2004}. Stars labelled with an asterisk are active M dwarfs. Top: UV and IR spectra up to 3~$\mu$m with linear axes. Bottom: UV spectra with logarithmic axes.}
     \label{figure:spectra}
\end{figure*}
\begin{figure}
   \resizebox{\hsize}{!}{\includegraphics{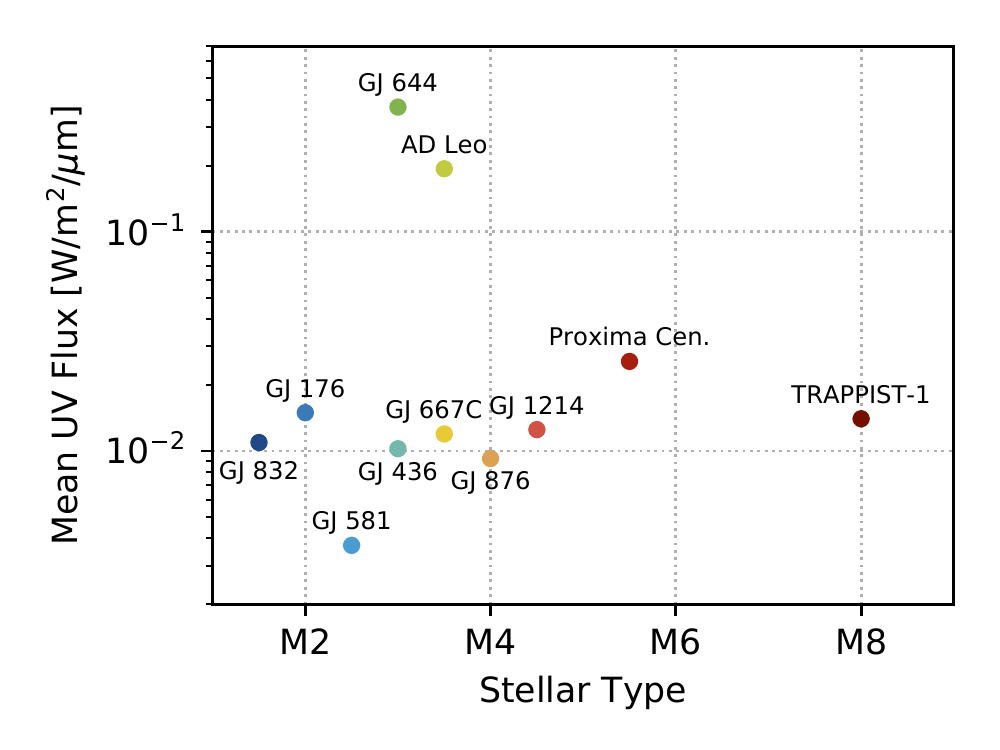}}
     \caption{Mean UV flux of each M dwarf between 170~nm and~240~nm. Mean values are calculated from the stellar spectra, scaled as in Fig~\ref{figure:spectra}.}
     \label{figure:uvcomp}
\end{figure}
An important aim of this study is to investigate the influence of different spectral energy distributions (SED) on the temperature and chemical composition profiles of an Earth-like planet in the HZ around an M dwarf. We used eight~observed M-dwarf spectra from the MUSCLES database\footnote{archive.stsci.edu/prepds/muscles/} \citep{france2016}, the observed spectra of AD Leo and GJ~644 from the VPL website\footnote{vpl.astro.washington.edu/spectra/stellar/mstar.htm} \citep{segura2005}, and the constructed stellar spectrum of TRAPPIST-1 with high UV activity using the method of \citet{rugheimer2015} and \citet{omalley2017}. We note that there is no observed spectrum of TRAPPIST-1 in the UV. The spectrum used in this study is assumed and could be very different from the real spectrum. The Mega-MUSCLES HST Treasury Survey is planning to observe TRAPPIST-1 to construct a representative spectrum of this low mass M dwarf \citep{froning2018}. \\
The spectra cover a range from M1.5 to M8 spectral type. For comparison we used the solar spectrum from \citet{gueymard2004}. Table \ref{table:stars} lists basic parameters of all stars used in this study. 
Figure \ref{figure:spectra} compares all M-dwarf spectra and the solar spectrum and Fig. \ref{figure:uvcomp} shows the mean UV flux of each M dwarf between 170 nm and~240~nm. In both plots the flux is scaled so that the surface temperature of the modelled planet reaches a value of 288.15~K. The scaling values are shown in Table \ref{table:planets}. Whereas the MUSCLES stars are classified as inactive \citep{france2016}, AD~Leo and GJ~644 are often cited as active M dwarfs \citep[e.g.][]{segura2005, reiners2009, leitzinger2010, rauer2011, rugheimer2015}. We label the active M dwarfs with an asterisk to separate them from non-active M dwarfs. Since the real TRAPPIST-1 spectrum is unknown, we cannot make any conclusions regarding the UV activity of this star. The spectrum we adopted for this study has a low mean UV flux, since we do not label TRAPPIST-1 as active.
From the MUSCLES database we used version 22 of the adapted panchromatic SED, binned to a constant resolution of 1~$\AA$. In low S/N regions the spectra were sampled down by the MUSCLES project to avoid negative fluxes by averaging negative flux bins with their neighbours until no negative flux bins remained. The spectra of GJ~1214, GJ~436 and Proxima Centauri include zero values for some wavelength bins in the UV. For these bins we linearly interpolated the flux with the two next non-zero neighbours to achieve a better estimate of the real flux. We applied the same downsampling procedure to avoid negative and non-zero flux to the spectra of AD Leo and GJ~644.\\
Since the MUSCLES spectra extend only up to a wavelength of 5.5~$\mu$m, we extended the available wavelength range with the NextGen\footnote{phoenix.ens-lyon.fr/Grids/NextGen/} spectra up to 971~$\mu$m \citep{hauschildt1999} to avoid overestimation of the flux when scaling to 1~solar constant. For the MUSCLES spectra we took the values of the effective temperature, log~$g$ and [Fe/H] from \citet{loyd2016}. Since the NextGen spectra have a fixed grid for effective temperature, log~$g$ and [Fe/H], we took the spectra with the most similar log~$g$ and [Fe/H] values and linear interpolated the temperature grid to the corresponding value. \\
The incoming stellar radiation is used up to 4.545~$\mu$m in the radiative transfer calculations. For a G-type star like the Sun $\sim$ 1~$\%$ of the incoming flux is neglected by the model because of this cut. For TRAPPIST-1 the flux which is not included at longer wavelengths by the model increases up to 4.6 $\%$ of the total incoming radiation. 
\begin{table*}
\centering
\caption{Parameters of the Earth around the Sun and Earth-like planets around M-dwarf host stars. The value $S/S_{\odot}$ is the ratio of the total stellar irradiance to the TSI to reach a surface temperature of 288.15~K. The transit duration was calculated via Eq.~\ref{eq:td}. We assume circular orbits for all planets. The transit depth, $\delta~=~\frac{R_\text{p}^2}{R_\text{s}^2}$, is calculated using the stellar radii from Table~\ref{table:stars} and the radius of the Earth. The equilibrium temperature (T$_\text{eq}$) was calculated using the corresponding planetary albedo ($\alpha_\text{p}$).}              
\label{table:planets}      
\centering                                     
\begin{tabular}{l r r r r r r r}          
\hline\hline                        
Host star       & $S/S_\odot$ & $a$ [AU]        & $P$ [d]       & $t_\text{D}$ [h]     & $\delta$ [ppm]        & $\alpha_\text{p}$     & $T_\text{eq}$ [K]\\ 
\hline 
Sun             & 1.000         & 1.000         & 365.3         & 13.10                 & 84.1                    & 0.201                 & 263.4 \\
GJ 832          & 0.886         & 0.205         & 50.4          & 4.28                  & 364.8                   & 0.117                 & 262.0 \\
GJ 176          & 0.915         & 0.192         & 45.9          & 3.92                  & 409.6                   & 0.095                 & 265.8 \\
GJ 581          & 0.920         & 0.114         & 25.8          & 2.47                  & 940.1                   & 0.086                 & 266.8 \\
GJ 436          & 0.918         & 0.166         & 34.8          & 3.45                  & 406.0                   & 0.086                 & 266.6 \\
GJ 644          & 0.888         & 0.242         & 67.6          & 6.81                  & 182.8                   & 0.096                 & 263.7 \\
AD Leo          & 0.878         & 0.143         & 30.4          & 3.02                  & 552.6                   & 0.094                 & 263.1 \\
GJ 667C         & 0.913         & 0.162         & 41.5          & 4.27                  & 397.2                   & 0.096                 & 265.5 \\
GJ 876          & 0.951         & 0.113         & 22.9          & 2.77                  & 594.5                   & 0.071                 & 270.1 \\
GJ 1214         & 0.958         & 0.068         & 15.6          & 1.78                  & 1887.9          & 0.068                 & 270.8 \\
Proxima Cen.    & 0.986         & 0.040         & 8.4           & 1.13                  & 4227.7          & 0.059                 & 273.4 \\
TRAPPIST-1      & 1.045         & 0.022         & 4.4           & 0.87                  & 6036.4          & 0.053                 & 277.9 \\

\hline 
\end{tabular}
\end{table*}
\subsection{Model scenarios}
We model Earth-like planets around M dwarfs with N$_2$-O$_2$ dominated atmospheric chemical compositions. We use 78 $\%$ nitrogen, 21 $\%$ oxygen, and a modern CO$_2$ volume mixing ratio of 355~ppm as starting conditions. While in previous studies it was assumed that these concentrations are isoprofiles which stay constant with altitude, in this study only the volume mixing ratios (vmr) at the surface are held to these values. The vmr within the atmosphere are allowed to vary (see \citealt{gebauer2017} for details). The chemistry module uses Earth-like boundary conditions to reproduce mean Earth atmospheric conditions. These are kept constant for each simulation and are further described in \citet{gebauer2017}. We use the Earth's radius and gravity and a surface pressure of 1~bar for all modelled planets. The temperature profile is calculated with the climate module depending on the chemical composition. The tropospheric water concentration is calculated via the assumption of an Earth-like relative humidity profile  \citep[see][]{manabe1967}. In contrast to \citet{rauer2011}, but following \citet{segura2005} we scale the stellar spectra from Fig. \ref{figure:spectra} so that we reproduce the Earth's surface temperature of 288.15~K, since the assumption of an Earth-like relative humidity may not be appropriate for higher temperatures \citep[see e.g.][]{leconte2013,godolt2015,godolt2016}. Table \ref{table:planets} shows the scaling values of the stellar insolation for each star.
We note that except the Earth, we do not model existing planets in this study. There are several studies investigating the potential atmosphere of Proxima Centauri~b \citep{kreidberg2016,turbet2016,meadows2018}, GJ~1214~b \citep{menou2011,berta2012,charnay2015}, GJ~436~b \citep{lewis2010,beaulieu2011} and the TRAPPIST-1 planets \citep{barstow_irwin2016,dewit2016,omalley2017}.

\subsection{Line-by-line spectral model}
The Generic Atmospheric Radiation Line-by-line Infrared Code (GARLIC) is used to calculate synthetic spectra between 0.4 $\mu$m and 12 $\mu$m \citep{schreier2014,schreier2018}. The GARLIC model is based on the Fortran 77 code MIRART-SQuIRRL \citep{schreier2001}, which has been used for radiative transfer modelling in previous studies like \citet{rauer2011} and \citet{hedelt2013}. The line parameters over the whole wavelength range are taken from the HITRAN 2016 database \citep{gordon2017}. Additionally the Clough-Kneizys-Davies continuum model \citep[CKD;][]{clough1989} and Rayleigh extinction are considered \citep{murphy1977,clough1989, sneep2005, marcq2011}. The transmission spectra are calculated using the temperature profile and the profiles of the 23 atmospheric species\footnote{OH, HO$_2$, H$_2$O$_2$, H$_2$CO, H$_2$O, H$_2$, O$_3$, CH$_4$, CO, N$_2$O, NO, NO$_2$, HNO$_3$, ClO, CH$_3$Cl, HOCl, HCl, ClONO$_2$, H$_2$S, SO$_2$, O$_2$, CO$_2$, N$_2$}, which HITRAN 2016 and our 1D climate-chemistry model have in common. We note that not all of the species are relevant for transmission spectroscopy of Earth-like atmospheres \citep[see e.g.][]{schreier2018}. \\
The radius of a planet with an atmosphere is wavelength dependent. This effect may be measured for example via transit spectroscopy. The difference between the geometric transit depth (without the contribution of the atmosphere) and the transit depth (considering the atmosphere) is called the effective height $h_\text{e}(\lambda)$. We simulate the transmission spectra $\mathcal{T}(\lambda,z)$ for $L$~=~64 adjacent tangential beams (corresponding to the 64 chemistry levels $z$) through the atmosphere. The effective height of each atmosphere is calculated using
\begin{equation} \label{eq:effhei}  
 h_\text{e}(\lambda) =  \int_{0}^{\infty} \Big(1-\mathcal{T}(\lambda,z)\Big)~dz.
\end{equation}

\subsection{Signal-to-noise ratio model}
\label{sec:snr_model}
To calculate the S/N for potential measurements of our synthetic Earth-like planets, we further developed the background-limited S/N model used by \citet{vonparis2011} and \citet{hedelt2013}. We added the readout noise contribution to the model and calculate the duty cycle (fraction of time spent on target) to better evaluate the detectability of atmospheric spectral features. The code of \citet{vonparis2011} and \citet{hedelt2013} is based on the photon noise model applied in \citet{rauer2011} as follows:
\begin{equation} \label{eq:star_signal}
 S/N = S/N_\text{s} \cdot \frac{f_\text{A}}{\sqrt{2}} =  \frac{F_\text{s} \cdot t_\text{int} \cdot f_\text{A}}{\sigma_\text{total}},
\end{equation} 
where S/N$_\text{s}$ is the stellar S/N. The stellar signal, $F_\text{s}$ [photons/s], is calculated with
\begin{equation}
 F_\text{s} = \frac{1}{N} \cdot \frac{R_\text{s}^2}{d^2} I_\text{s} \cdot A \cdot q \cdot \Delta \lambda,
\end{equation}
where $R_s$ is the stellar radius, $I_\text{s}$ the spectral energy flux [W/m$^2$/$\mu$m], $d$ the distance of the star to the telescope, and $A$ the telescope area. The flux is divided by $N = h\frac{c}{\lambda}$ ($h$, Planck constant, $c$, speed of light) to convert into number of photons. In contrast to \citet{rauer2011} we consider a wavelength dependent total throughput, $q$, of the corresponding instrument. The bandwidth, $\Delta \lambda$ (resolving power, $R = \frac{\lambda}{\Delta \lambda}$), is dependent on the filter and disperser of the instrument in question.
The additional transit depth due to the atmosphere of the planet, $f_\text{A}$, is calculated using $R_s$, the planetary radius, $R_\text{p}$ and $h_\text{e}$ (Eq. \ref{eq:effhei}), i.e.
\begin{equation} \label{eq:atm_td}
 f_\text{A} = \frac{(R_\text{p} + h_\text{e})^2}{R_\text{s}^2} - \frac{R_\text{p}^2}{R_\text{s}^2},
\end{equation} 
The total noise contribution, $\sigma_\text{total}$, includes photon noise, $\sigma_\text{photon}$, zodiacal noise, $\sigma_\text{zodi}$, thermal noise $\sigma_\text{thermal}$, dark noise, $\sigma_\text{dark}$, and readout noise, $\sigma_\text{read}$, i.e.
\begin{equation}
 \sigma_\text{total} = \sqrt{2 \cdot (\sigma_\text{photon}^2 + \sigma_\text{zodi}^2 + \sigma_\text{thermal}^2 + \sigma_\text{dark}^2 + \sigma_\text{read}^2)}.
\end{equation}
The additional factor $\sqrt{2}$ is included to consider the planetary signal as a result of the difference between in and out of transit.
The individual noise sources are described in more detail in \citet{vonparis2011}. In constrast to \citet{vonparis2011} and \citet{hedelt2013} we calculate the smallest number of pixels, $N_\text{pixel}$, needed to collect at least 99\% of the full stellar signal. This assumption significantly decreases the contribution due to zodiacal, thermal, dark, and readout noise by neglecting contamination of background noise dominated pixels.
The number of photons per pixel is given by the point spread function (PSF) of the corresponding instrument, filter, and disperser. 
We use the pixel of the PSF with the highest fraction of the total signal to calculate the
exposure time, $t_\text{exp}$, which is the longest time of photon integration before the collector is saturated. Possible exposure times are usually predefined by the instrument configuration. The minimum exposure time defines the saturation limits of the instrument. 
The number of exposures, $N_\text{exp}$, during a single transit is calculated via
\begin{equation}
 N_\text{exp} = \Bigl\lfloor\frac{t_\text{D}}{t_\text{exp} + t_\text{read}}\Bigr\rfloor,
\end{equation}
where $t_\text{read}$ is the readout time of the instrument configuration and $t_\text{D}$ is the transit duration. The value $N_\text{exp}$ is rounded down to the next integer. \\
We calculate $t_\text{D}$ via 
\begin{equation} \label{eq:td}
 t_\text{D}~=~P/\pi~\cdot~\arcsin\Big(\sqrt{(R_\text{s}~+~R_\text{p})^2~-~b^2}/{a}\Big),
\end{equation}
with $b~=~a/R_\text{s}~\cos(i)$. The inclination, $i$, was assumed to be 90$^\circ$ for all planets.
The orbital distance, $a$, was calculated by $a~=~\sqrt{\Big((R/R_{\odot})^2 \cdot (T_\text{eff}/T_\text{eff~$\odot$})^4\Big)~/~\Big(S/S_\text{$\odot$}\Big)}$ and the orbital period, $P$, was calculated using Kepler's third law.
The full integration time of one transit, $t_\text{int}$ is considered for the noise calculation, as well as for the calculation of the number of photons collected from the star during one transit ($F_\text{s}~\cdot~t_\text{int}$). The value $t_\text{int}$ is calculated via
\begin{equation}
 t_\text{int} = t_\text{D} - N_\text{exp} \cdot t_\text{read}.
\end{equation}
The instrumental readout noise, RON, is usually given per pixel and per exposure. To calculate the full readout noise during one transit we use
\begin{equation}
 \sigma_\text{read} = N_\text{exp} \cdot N_\text{pixel} \cdot \text{RON}.
\end{equation}


\section{Results and discussion}
\subsection{Atmospheric profiles}
\begin{figure}
   \resizebox{\hsize}{!}{\includegraphics[width=8cm]{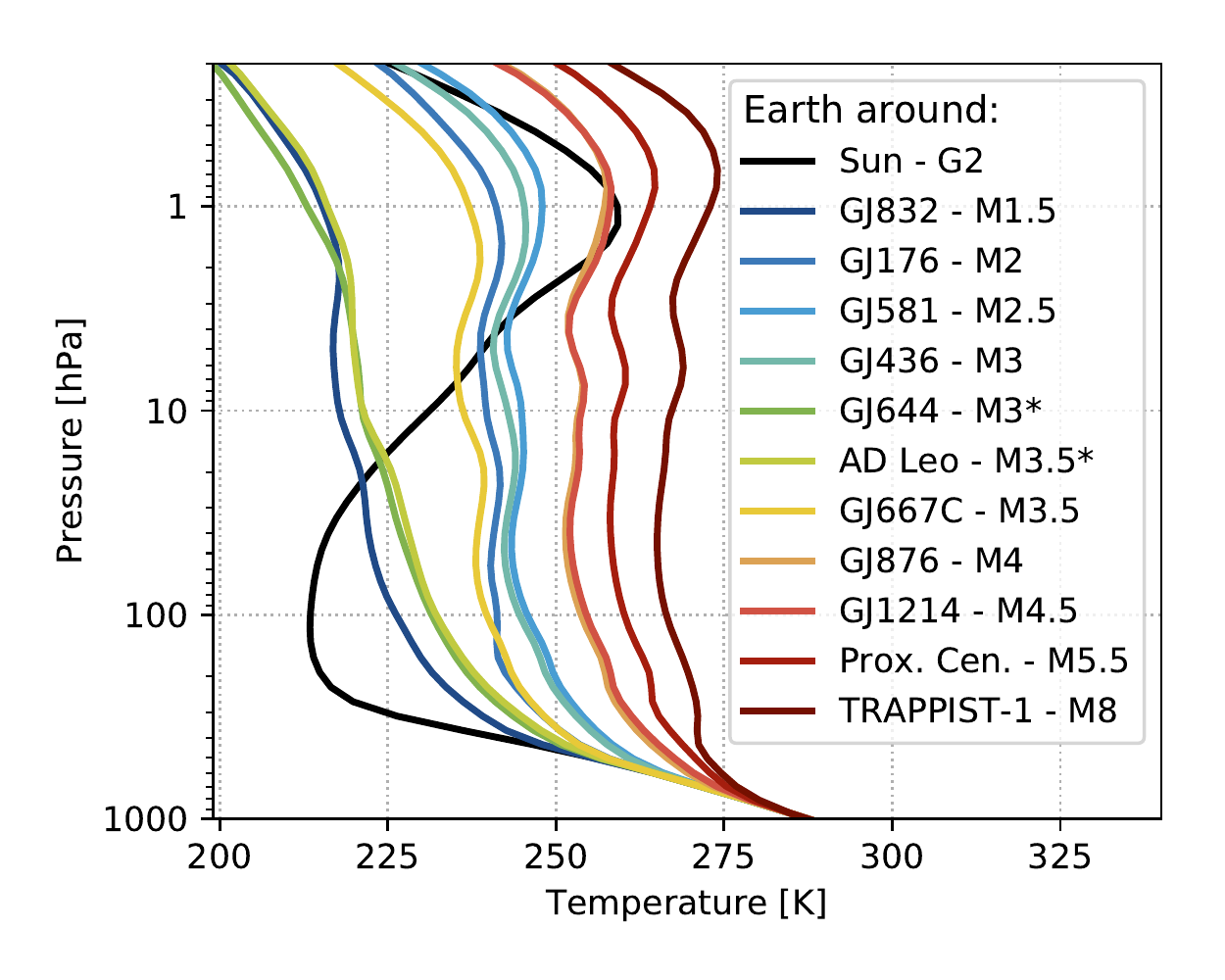}}
     \caption{Influence of various M-dwarf SEDs on the temperature profile~[K] of an Earth-like planet. The stellar spectra of Fig. \ref{figure:spectra} are scaled to reproduce a surface temperature of 288.15~K (see Table \ref{table:planets}). Each coloured line represents the temperature profile of the modelled hypothetical planet orbiting a different host star. Weak UV emission of M dwarfs leads to low O$_3$ heating in the middle atmosphere and therefore to a reduced or missing temperature inversion. Enhanced CH$_4$ heating increases the temperature in the middle atmosphere of Earth-like planets around late-type M dwarfs by up to 60~K. }
     \label{figure:temperature}
\end{figure}

\begin{figure*}
\centering
   \includegraphics[width=17cm]{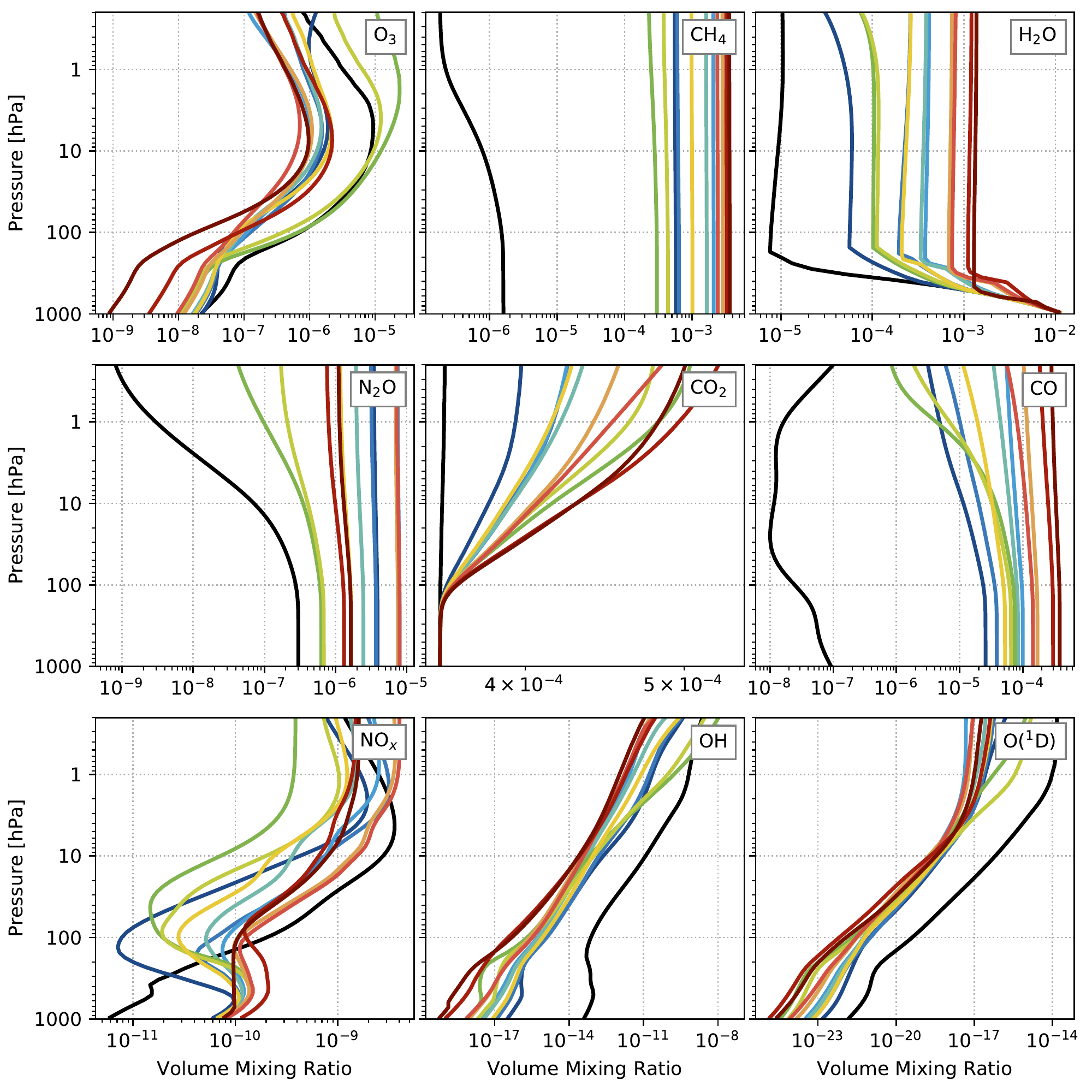}
     \caption{Influence of various M-dwarf SEDs on some selected chemical atmospheric profiles of an Earth-like planet. The legend is the same as for Fig. \ref{figure:temperature}.}
     \label{figure:chemical_profiles}
\end{figure*}

Figures \ref{figure:temperature} and \ref{figure:chemical_profiles} show the results from the climate-chemistry model using the M dwarf and the solar spectra as incoming flux shown in Fig. \ref{figure:spectra}. Compared to previous studies such as \citet{rauer2011}, \citet{grenfell2014}, and \citet{keles2018} we use a further developed model and a set of active and inactive observed and modelled M-dwarf SEDs. 

\subsubsection{Temperature profile}
Figure~\ref{figure:temperature} shows temperature profiles for the hypothetical Earth-like planets around their corresponding host stars. All planets are placed at a distance to their host star which results in a surface temperature of 288.15~K. Table~\ref{table:planets} shows the scaling factor of the total solar irradiance (TSI) of each planet. Without scaling the spectra, Earth-like planets around most M dwarfs would show an increased surface temperature \citep[see e.g.][]{rauer2011, rugheimer2015}. In part, this is due to the shift of the emission maximum of M dwarfs to the near-IR, which leads to increased absorption of CO$_2$ and H$_2$O and decreased scattering at UV and visible wavelengths in the planetary atmosphere \citep[see e.g.][]{kasting1993}. \\
This effect can be intensified by an increased greenhouse heating due to the increased amount of CH$_4$ and H$_2$O in the troposphere of Earth-like planets around M dwarfs \citep[see e.g.][]{rauer2011,grenfell2013,rugheimer2015}. Most M-dwarf cases require decreased TSI to reach a surface temperature of 288.15~K. However, above a certain CH$_4$ concentration, the surface temperature decreases with the same insolation. This suggests that most of the stellar irradiation is absorbed in the stratosphere, resulting in a smaller increase of the surface temperature due to reduced stellar irradiation reaching the troposphere \citep[see also][]{grenfell2013}. For M1 to M3 type stars the TSI needs to be decreased by about 8-12$\%$. Mid M dwarfs show this saturation in greenhouse heating, resulting in a minor downscaling of 1-5$\%$. For the synthetic planet orbiting the late M-dwarf TRAPPIST-1, the stellar flux needs to be increased by 5$\%$ to reproduce Earth's surface temperature. Hence, the chemistry feedback counteracts the effect of the SED, which may shift the inner edge of the HZ towards shorter distances to the star for late-type M-dwarf planets compared to climate-only simulations such as \citet{kopparapu2013}. Neglecting the composition change, hence assuming Earth's composition and carrying out climate only calculations suggest that the TSI would need to be decreased by 8-12$\%$ for all Earth-like planets around M dwarfs including TRAPPIST-1 and Proxima~Centauri (not shown). Hence the chemical feedback tends to cool the planets around late-type M dwarfs. This result is consistent with the study of \citet{ramirez2018} who found that for planets around late M dwarfs a CH$_4$ anti-greenhouse effect can shrink the HZ by about 20\%.
\\
Our simulations do not include effects on the radiative transfer due to clouds and hazes. This decreases the equilibrium temperature of the Earth due to the lower planetary albedo (see Table~\ref{table:planets}). The consideration of clouds can decrease or increase the surface temperature depending on the type and height of the clouds \citep[see e.g.][]{kitzmann2010}. Including thick hazes would likely decrease the surface temperature by a few degrees \citep[see e.g.][]{arney2016,arney2017}. Even for climate models of the Earth it is still challenging to simulate the effect of clouds and aerosols consistently \citep[see e.g.][]{ipcc2014}. Hence, we do not consider clouds and hazes for this model study. Effects which would lead to a different surface temperature could be corrected by up- or downscaling the stellar flux to obtain a surface temperature of 288.15~K. \\
In the middle atmosphere (stratosphere and mesosphere up to 0.01~hPa) heating due to O$_3$ absorption is decreased owing to weak UV emission of M dwarfs \citep[see e.g.][]{segura2005,grenfell2013,grenfell2014}. Therefore, no temperature inversion is present in the stratosphere, as found by \citet{segura2005}. For active and inactive early M dwarfs the temperature maximum in the middle atmosphere around 1~hPa is decreased compared to the Earth around the Sun. The enhanced heating rates due to CH$_4$ absorption of stellar irradiation increase the temperature in the middle atmosphere by up to 60~K for late-type M-dwarf planets compared to early-type M-dwarf planets \citep[see also][]{rauer2011,grenfell2014}. Increased temperatures from about 500~hPa up to 1~hPa result in an almost isothermal vertical gradient, most pronounced for late M-dwarf planets. For emission spectroscopy the weak temperature contrast between the surface and the atmosphere reduces the spectral features, resulting in low S/N as discussed in \citet{rauer2011}. 

\subsubsection{Responses of O$_3$}
The production of O$_3$  in the middle atmosphere is sensitive to the UV radiation. In the Schumann-Runge bands and Herzberg continuum (from about 170 nm to 240~nm) molecular oxygen (O$_2$) is split into atomic oxygen in the ground state (O$^3$P), which reacts with O$_2$ to form O$_3$. The destruction of O$_3$ is mainly driven by absorption in the Hartley (200 nm~-~310~nm), Huggins (310~nm -~400~nm), and Chappuis (400~nm ~-~850~nm) bands. Hydrogen oxides, HO$_x$ (OH, HO$_2$, and HO$_3$), and nitrogen oxides, NO$_x$ (NO, NO$_2$, and NO$_3$), destroy O$_3$ via catalytic loss cycles \citep[e.g.][]{brasseur2006}. All planets around inactive host stars from the MUSCLES database and TRAPPIST-1  have a reduced O$_2$ photolysis in the Schumann-Runge bands and Herzberg continuum compared to the Earth around the Sun case \citep[see Fig. \ref{figure:uvcomp} and ][]{rugheimer2015}. The decreased destruction of O$_3$ by photolysis at wavelengths longer than $\sim$200~nm \citep[see Fig. \ref{figure:spectra} and][]{grenfell2014} can not compensate the decreased production, resulting in lower concentrations of O$_3$ throughout the atmosphere (see Fig. \ref{figure:chemical_profiles}). Earth-like planets around active M dwarfs (AD~Leo and GJ~644) show strong O$_2$ photolysis in the Schumann-Runge bands and Herzberg continuum (see Fig. \ref{figure:uvcomp}), which results in an increase of O$_3$ concentrations compared to planets around inactive M dwarfs \citep[see also][]{grenfell2014}.

\subsubsection{Responses of CH$_4$}
For an Earth-like atmosphere the destruction of CH$_4$ is mainly driven by the reaction with the hydroxyl radical, OH, (CH$_4$~+~OH~$\rightarrow$~CH$_3$~+~H$_2$O). The amount of OH in the atmosphere is closely associated with UV radiation via the source reaction with excited oxygen (O$^1$D): H$_2$O~+~O$^1$D~$\rightarrow$~2OH. O$^1$D is mainly produced by O$_3$ photolysis (see e.g. \citealt{grenfell2007} and \citealt{grenfell2013} for further details). As in previous studies \citep[e.g.][]{rauer2011} we assume a constant CH$_4$ bioflux for all our simulations. No in situ chemical production of CH$_4$ in the atmosphere is included in the model. The effect of different CH$_4$ biomass emissions on planets around M dwarfs was investigated by for example \citet{grenfell2014}.
Compared to the Earth around the Sun, the UV photolysis of O$_3$ is reduced for all Earth-like planets orbiting M dwarfs \citep[see][]{grenfell2013}. This leads to reduced O$^1$D and therefore OH concentrations, except for active M-dwarf planets, where OH is increased in the upper middle atmosphere (Fig.~\ref{figure:chemical_profiles}). The reduced OH concentrations decrease the destruction of CH$_4$ and lead to enhanced CH$_4$ abundances \citep{segura2005,grenfell2013,grenfell2014,rugheimer2015}. The planets around the inactive MUSCLES host stars show  CH$_4$ concentrations comparable to the results of \citet{rugheimer2015}. The active host stars have higher far UV radiation, which leads to more O$_3$ production, increased OH concentrations in the upper middle atmosphere, and more CH$_4$ destruction in the atmosphere of Earth-like planets compared to planets orbiting inactive host stars.  \\
All our simulated planets around M dwarfs feature a CH$_4$/CO$_2$ ratio higher than 0.2, for which previous model studies suggest that hydrocarbon haze formation is favoured \citep[see][]{trainer2004,arney2016,arney2017}. Our model does not include hydrocarbon haze production or the impact of haze on the radiatve transfer. \citet{arney2017} found that the surface temperature of an Earth-like planet around AD~Leo would decrease by 7~K when the effect of hazes is considered. Such a decrease of the surface temperature by hazes would require an increase in stellar flux to obtain a surface temperature of 288.15 K. On the one hand haze production would lead to an CH$_4$ sink, on the other hand the lower UV radiation due to UV absorption by organic haze would decrease the CH$_4$ photolysis. The overall effect upon the resulting CH$_4$ profile is difficult to estimate. Hence, we refer to \citet{arney2016} and \citet{arney2017} who investigated the impact of organic haze on Earth-like atmospheres. \\
We based our simulations on the assumption of a constant CH$_4$ bioflux for all planets. \citet{segura2005} and \citet{rugheimer2015} reduced the surface flux of CH$_4$ or set a constant surface mixing ratio to limit the building up of CH$_4$ due to the reduced destruction for Earth-like planets around late M dwarfs. Although the biosphere of an Earth-like planet around M dwarfs is likely to be different to the Earth, we cannot exclude that biofluxes might be similar.

\subsubsection{Responses of H$_2$O}
In the middle atmosphere the increased CH$_4$ abundances lead to more H$_2$O production via CH$_4$ oxidation for Earth-like planets around M dwarfs compared to the Earth-Sun case \citep[e.g.][]{segura2005,rauer2011}. In the troposphere the H$_2$O content is related to the temperature profile and is calculated with the assumption of a relative humidity profile. Although all planets are placed at the distance to their host star where the surface temperature reaches 288.15 K, the H$_2$O content in the upper troposphere is increased for M-dwarf planets and is close to the mixing ratio thought to lead to water loss. In this case this high mixing ratio is caused by CH$_4$ oxidation. The temperatures in the troposphere are increased for all Earth-like M-dwarf planets compared to the Earth. This allows the atmosphere to sustain more H$_2$O in the troposphere, even if the same H$_2$O vmr is obtained at the surface. 
\label{sec:stellar_snr}
\begin{figure*}
\centering
   \includegraphics[width=17cm]{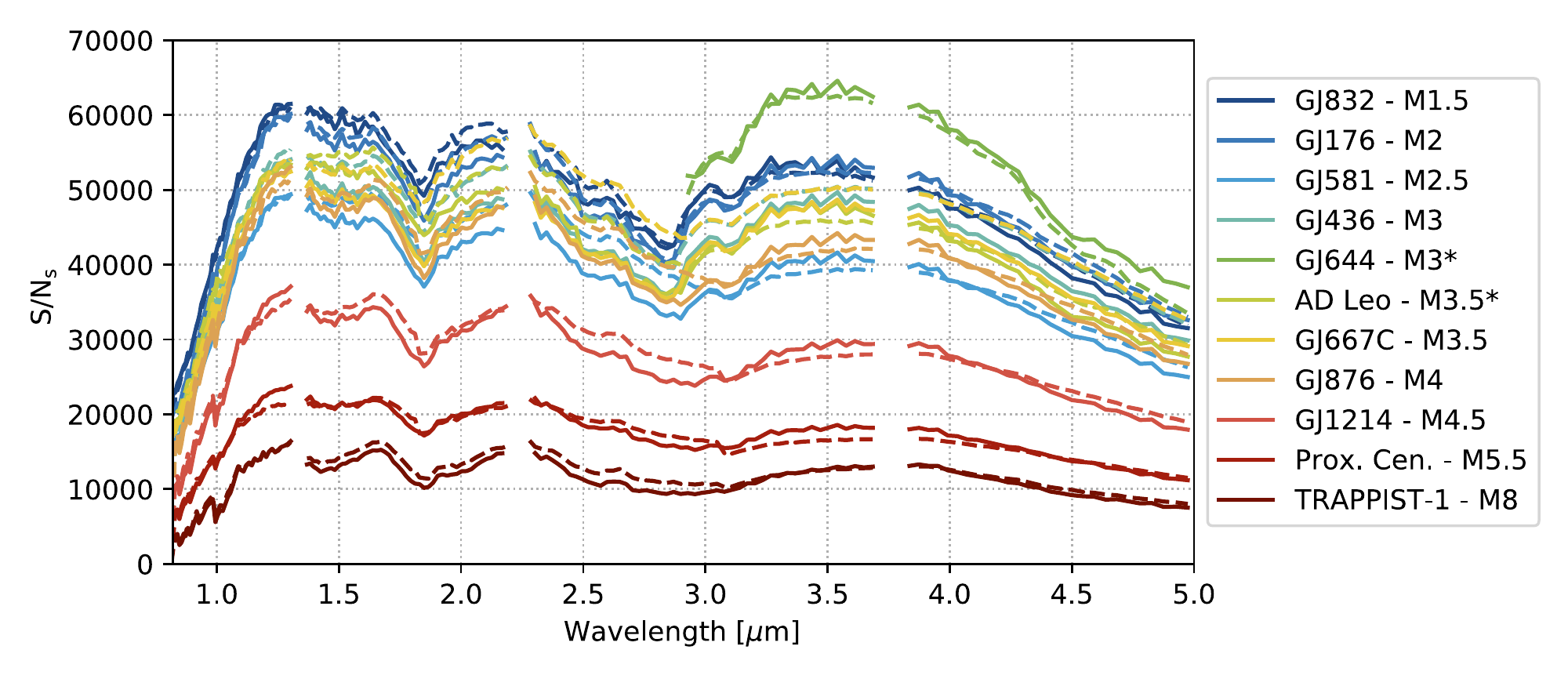}
     \caption{Stellar signal-to-noise ratio over 1 h integration time of each M dwarf at 10~pc, binned to a resolution of 100. All NIRSpec filters and high resolution disperser are combined. Solid lines are calculated using the method described in Sec. \ref{sec:snr_model} and dashed lines are calculated using the JWST  ETC, Pandeia \citep{pontoppidan2016}.}
     \label{figure:comp2pan}
\end{figure*}

\subsubsection{Responses of N$_2$O}
On Earth nitrous oxide (N$_2$O) sources are mainly surface biomass emissions. Only weak chemical sources in the atmosphere further increase the amount of N$_2$O in the atmosphere \citep[see e.g.][]{grenfell2013}. We assume constant N$_2$O biomass emissions for all our simulations bases on modern Earth \citep[see e.g.][]{grenfell2011,grenfell2014}. The loss processes of N$_2$O are dominated by photolysis in the UV below 240~nm in the middle atmosphere (via N$_2$O~+~h$\nu$~$\rightarrow$~N$_2$~+~O$^3$P). We use the cross sections of \citet{selwyn1977}, which cover the wavelength range from 173 nm to 240~nm and peak for room temperature at around 180~nm. For Earth-like planets around M dwarfs and the Earth only $\sim$ 5-10$\%$ of N$_2$O is destroyed via catalytic reaction with O$^1$D \citep[see e.g.][]{grenfell2013}. The resulting N$_2$O is therefore closely related to the SED around 180~nm (see Fig.~\ref{figure:uvcomp}). The SED in the far UV range is dominated by the activity of the M dwarf and has a lower dependence on the spectral type \citep{france2013}. Since active M dwarfs emit strongly in the far UV, planets orbiting active M dwarfs show reduced N$_2$O abundances in comparison to planets around inactive M dwarfs. Similar to \citet{rugheimer2015} we do not find a dependence of the N$_2$O abundances on the type of the M dwarf using observed stellar spectra. The amount of N$_2$O in the atmosphere of Earth-like planets is mainly dependent on the incoming radiation in the wavelength range between 170 nm and~240~nm of the host star (see Fig.~\ref{figure:uvcomp}). For modelled spectra \citet{rugheimer2015} show that N$_2$O increases for Earth-like planets around late-type M dwarf.

\subsubsection{Responses of CO$_2$}
The CO$_2$ production is mainly driven by the reaction CO~+~OH~$\rightarrow$ ~CO$_2$~+~H. The increased CH$_4$ oxidation in the middle atmosphere leads to an increase in CO for Earth-like planets around M dwarfs compared to the Earth-Sun case (Fig.~\ref{figure:chemical_profiles}). CO$_2$ is mainly destroyed by photolysis in the far UV. At wavelengths longer than 120~nm the photolysis of CO$_2$ is strongest between 130~nm and~160~nm \citep[e.g.][]{huestis2011}. Despite the strong emission of AD~Leo in the far UV, \citet{gebauer2017} found a 40~\% increase of CO$_2$ abundances in the upper middle atmosphere of an Earth-like planet around AD~Leo due to an increase of OH and CO at these altitudes compared to the Earth around the Sun. Even for lower abundances of OH in the upper middle atmosphere, we find that CO$_2$ increases in the middle atmosphere for all M-dwarf planets because of the large amounts of CO.
The enhancement of the CO$_2$ production was not taken into account in previous studies, such as \citet{rauer2011} and \citet{grenfell2014}, where CO$_2$ was kept constant. At the surface all scenarios show a CO$_2$ volume mixing ratio of 355 ppm.

\subsection{Stellar S/N}
Just a fraction of the measured stellar signal of the order of a few parts per million (ppm) is blocked by the atmosphere of a terrestrial planet during a transit. Hence, to calculate the S/N of the atmosphere of the planet, we first need to know the stellar S/N (S/N$_\text{s}$). The radius of the star and the distance to the star are also important to determine the stellar signal in addition to the spectral type,  (Eq. \ref{eq:star_signal}). Figure~\ref{figure:comp2pan} shows the S/N$_\text{s}$ for all M dwarfs at 10~pc, observed for an integration time of 1 hour, using all high resolution JWST NIRSpec filters (G140H/F070LP, G140H/F100LP, G235H/F170LP, G395H/F290LP). The specifications used to calculate the S/N using the NIRSpec filters can be found in Appendix \ref{sec:appendix_nirspec}. We binned the S/N to a constant resolution of 100 over the entire wavelength range ($\Delta\lambda$ = 10~nm at 1~$\mu$m and $\Delta\lambda$ = 50~nm at 5~$\mu$m). We chose the highest S/N for all overlapping bins of two filters. We note that in practise it is not possible to observe several NIRSpec filters at the same time. Gaps in the wavelength coverage at $\sim$ 1.3~$\mu$m, $\sim$ 2.2~$\mu$m, and $\sim$ 3.8~$\mu$m result from a physical gap between the detectors\footnote{www.cosmos.esa.int/web/jwst-nirspec/bots}.\\
At 10~pc, all M dwarfs except GJ~644 are observable over the entire wavelength range using high resolution spectroscopy (HRS). For the low wavelength filters GJ~644 reaches the saturation limit at around 11.5~pc. The Sun, placed at 10 pc, is not observable at any wavelength using NIRSpec due to saturation of the detector. The SED of M dwarfs peaks between 0.7~$\mu$m (early M dwarfs) and 1.1~$\mu$m (late M dwarfs). Because of the broader bandwidth with increasing wavelengths the S/N does not decrease significantly towards longer wavelengths. \citet{nielsen2016} showed that the S/N for GJ~1214 reduces by a factor of 3 from 1.5~$\mu$m to 4~$\mu$m when using a constant bandwidth. The early M dwarfs are near the saturation limit, which results in a significant reduction of the S/N due to changes in the detector duty cycle (with shorter exposures) and additional readout noise \citep[see also][]{nielsen2016}. Early-type M dwarfs have larger S/N than late-type M dwarfs at the same distance owing to the combined effect of effective temperature and stellar radius.
\\
The solid lines of Fig.~\ref{figure:comp2pan} show the S/N calculated using the method described in Sec.~\ref{sec:snr_model}. Our S/N model can be applied to any space and ground-based telescope. In comparison to the previous S/N model version of \citet{hedelt2013} we added the readout noise contribution and saturation limits to the code. We plan to further develop this model and include exozodiacal dust contamination and the influence of star spots on the S/N of transmission and emission spectroscopy. \\
To verify our S/N model we compared the calculations to results from the JWST Exposure Time Calculator (ETC), Pandeia \citep{pontoppidan2016}. The S/N calculated with Pandeia are shown as dashed lines in Fig.~\ref{figure:comp2pan}. For all stars and wavelengths we find that our calculations are in good agreement with the S/N calculated with Pandeia. For both calculations GJ~644 is just observable using G395H/F290LP. This confirms that the saturation limits of the NIRSpec HRS are also in agreement.

\subsection{Transmission spectra}
\label{sec:transmission_spectra}
\begin{figure*}
\centering
   \includegraphics[width=17cm]{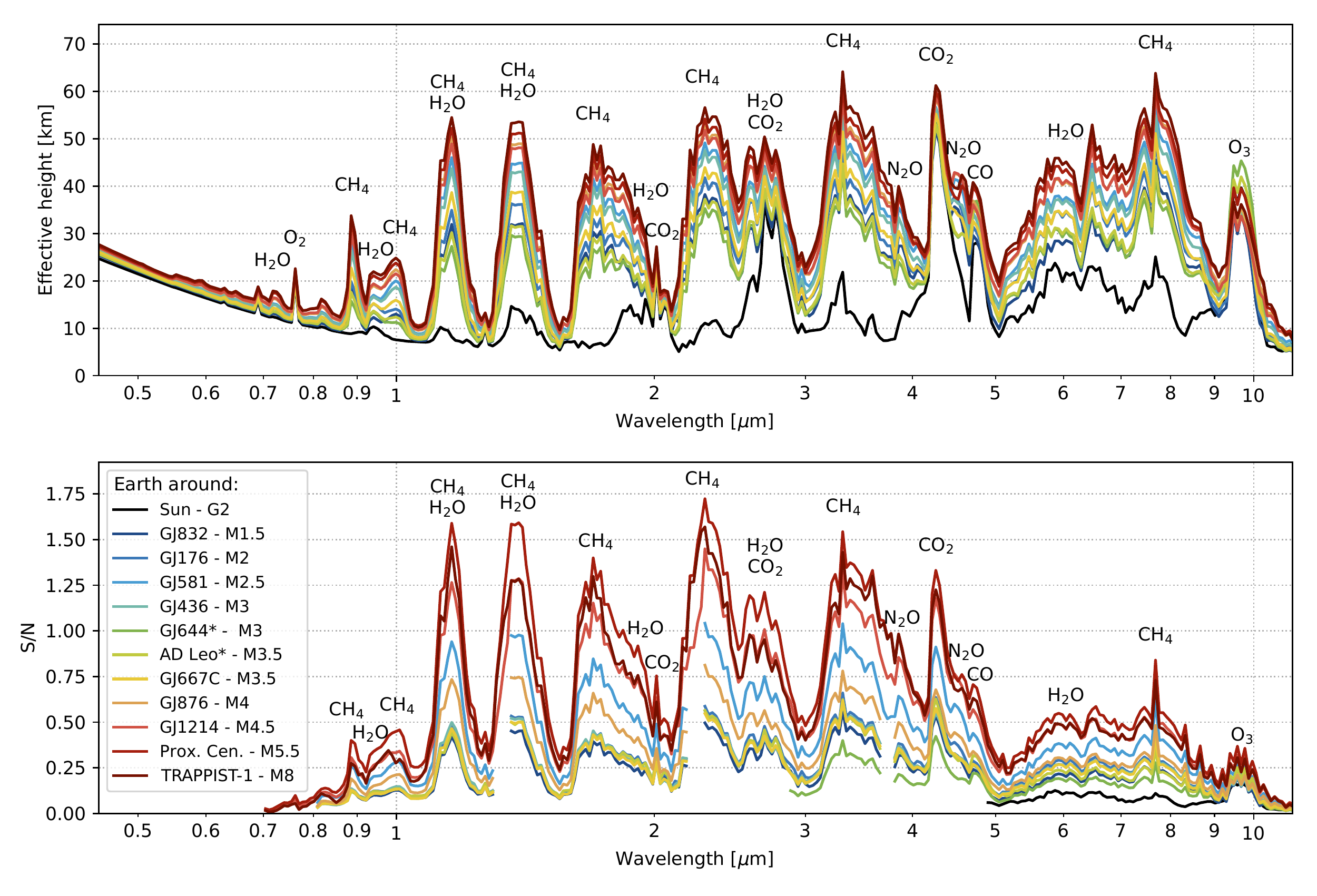}
     \caption{Top: Effective height of all hypothetical Earth-like planets orbiting M dwarfs for a fixed resolving power of 100. This data is available at \href{http://cdsarc.u-strasbg.fr/viz-bin/qcat?J/A+A/624/A49}{cdsarc.u-strasbg.fr/viz-bin/qcat?J/A+A/624/A49}. Bottom: combined S/N using all high or medium resolution NIRSpec and MIRI filters up to 11~$\mu$m for all M-dwarf planets at 10 parsecs integrated over a single transit and a resolving power of 100.}
     \label{figure:transmission_snr}
\end{figure*}

The top panel of Fig. \ref{figure:transmission_snr} shows the transmission spectra of each simulated hypothetical Earth-like planet around an M dwarf. The transmission spectra are shown as the wavelength dependent effective height (see Eq. \ref{eq:effhei}). We binned the line-by-line effective height of each planet to a resolving power of $R = \frac{\lambda}{\Delta\lambda} = 100$ over the entire wavelength range. 
We calculated the effective height of the Earth around the Sun for comparison. Dominant absorption features are H$_2$O at 1.4~$\mu$m and 6.3~$\mu$m, CO$_2$ at 2.8~$\mu$m, and 4.3~$\mu$m and O$_3$ at 9.6~$\mu$m, comparable to observed and modelled transmission spectra of the Earth \citep{palle2009,kaltenegger2009, vidal2010, betremieux2013, yan2015, schreier2018}. \\
We do not consider clouds and hazes when calculating the transmissivity of the atmosphere. The consideration of clouds would essentially limit the minimum effective height to the height of the cloud deck \citep[e.g.][]{benneke2013,betremieux2014, Betremieux2017}. For the Earth the cloud layer is low and therefore the extent of the absorption feature is affected by only a few kilometres \citep[e.g.][]{schreier2018}. The consideration of hazes in the atmosphere would effect the spectral appearance of the planet \citep[see e.g.][]{kempton2011,kreidberg2014}. For Earth-like planets hazes affect the plantary spectra mostly at wavelengths below about 3.2~$\mu$m \citep[see][]{arney2016,arney2017}. For hazy atmospheres the transit depth is increased between the absorption features of, for example CH$_4$, H$_2$O and O$_2$ (atmospheric windows), leading to a lower separability of the individual spectral bands. \\
For Earth-like planets around M dwarfs the amount of H$_2$O and CH$_4$ in the middle atmosphere is increased by several orders of magnitude (Fig. \ref{figure:chemical_profiles}). Since CH$_4$ and H$_2$O are strong absorbers in the near-IR, increased abundances result in a decreased transmissivity and larger effective heights at most wavelengths compared to the Earth around the Sun \citep[see also][]{rauer2011}. In the middle atmosphere CH$_4$ is more abundant than H$_2$O for Earth-like planets around M dwarfs. Hence, the spectral features at 1.1~$\mu$m and 1.4~$\mu$m are dominated by CH$_4$ absorption as also shown by \citet{barstow2016}. Larger temperatures in the middle atmosphere of planets around late M dwarfs (Fig. \ref{figure:temperature}) lead to an expansion of the atmosphere by more than 10~km compared to planets around early M dwarfs (Table~\ref{table:atmosphere_H_T}). This, together with enhanced absorption of CH$_4$ and H$_2$O, results in an overall increased effective height for mid to late M-dwarf planets.
Most spectral features of Earth-like planets around inactive M dwarfs are larger as a consequence of increased concentrations of CH$_4$, H$_2$O, and N$_2$O compared to planets around active M dwarfs. Just the spectral feature of O$_3$ is largest for active M-dwarf planets because of the increased O$_2$ photolysis in the UV, resulting in an enhanced O$_3$ layer. \\
Most other studies have used existing planets and have applied assumed atmospheres to spectral models \citep{shabram2011,barstow2015,barstow2016,barstow_irwin2016,morley2017}. In comparison to these studies the calculated concentrations of most biosignatures and related compounds are increased owing to the consideration of the chemical response of the atmosphere to M-dwarf spectra. Hence, the effective height of the atmosphere in our study is larger at most wavelengths and the detectability of most spectral features is increased. 
\begin{table}
\centering
\caption{Height, $H$, and temperature, $T$, at 100~hPa and at 0.1~hPa for all simulated atmospheres of Earth-like planets around M dwarfs and the Earth. Earth-like planets around late M dwarfs show an expansion of the atmosphere by more than 10~km compared to planets around early M dwarfs and the Earth.}              
\label{table:atmosphere_H_T}      
\centering                                     
\begin{tabular}{l r r r r}          
\hline\hline  
                &\multicolumn{2}{c}{100 hPa}    & \multicolumn{2}{c}{0.1 hPa}    \\
Host star       &       $H$ [km]        &       $T$ [K] &       $H$ [km]        &       $T$ [K]     \\
\hline 
Sun             &       16.2    &       213.5   &       63.3    &       199.0   \\
GJ 832          &       16.7    &       226.2   &       59.9    &       177.5   \\
GJ 176          &       17.2    &       241.2   &       64.9    &       201.1   \\
GJ 581          &       17.5    &       245.7   &       66.2    &       208.2   \\
GJ 436          &       17.4    &       244.6   &       65.7    &       205.2   \\
GJ 644          &       16.9    &       231.5   &       60.4    &       179.5   \\
AD Leo          &       17.0    &       232.2   &       60.8    &       182.4   \\
GJ 667C         &       17.2    &       239.7   &       64.2    &       198.4   \\
GJ 876          &       17.9    &       253.7   &       68.4    &       217.0   \\
GJ 1214         &       17.9    &       254.5   &       68.5    &       216.0   \\
Proxima Cen.    &       18.2    &       260.2   &       70.0    &       224.1   \\
TRAPPIST-1      &       18.4    &       266.4   &       72.0    &       235.1   \\
\hline 
\end{tabular}
\end{table}

\begin{figure}
   \resizebox{\hsize}{!}{\includegraphics{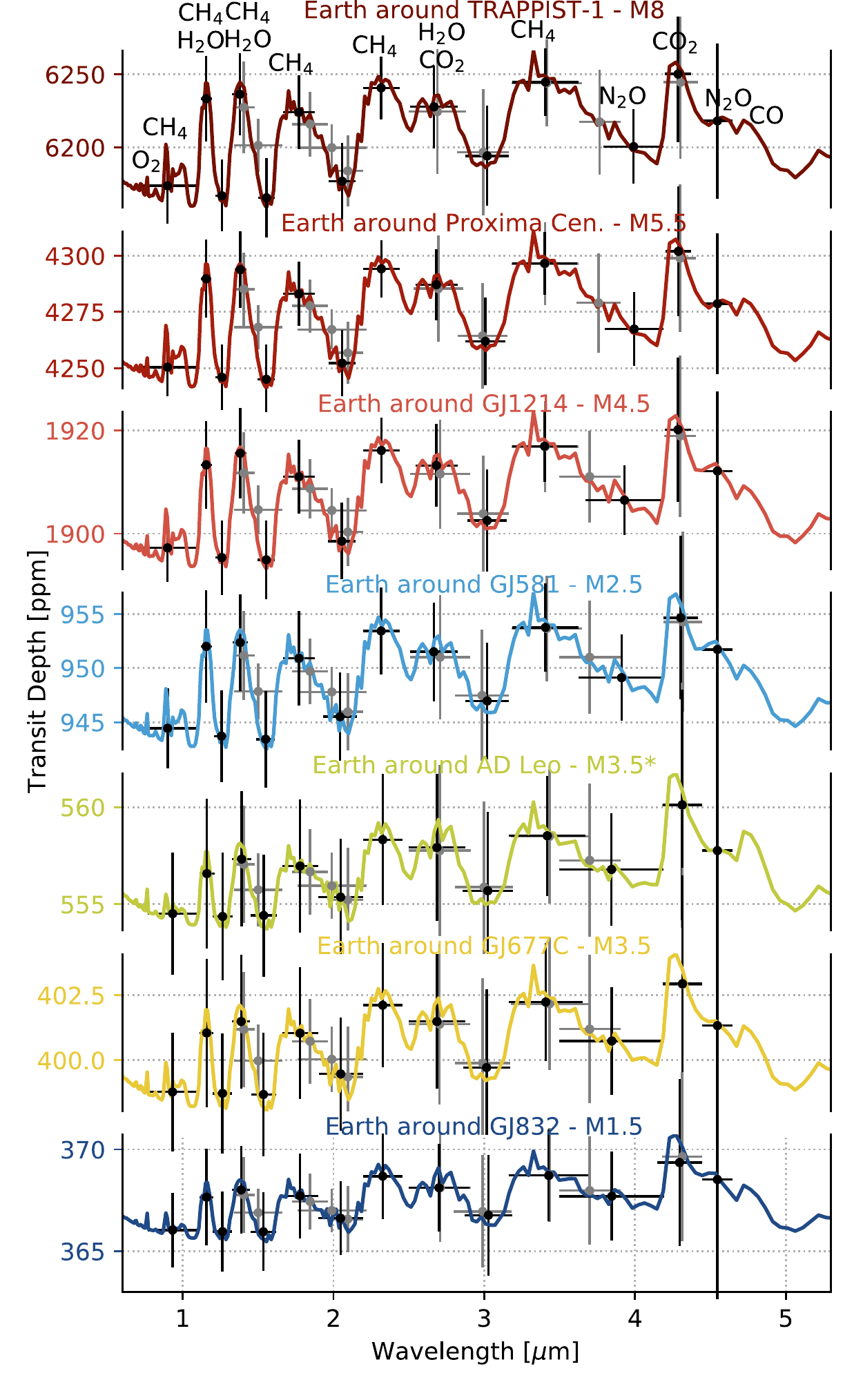}}
     \caption{Coloured lines show the transit depth of all hypothetical Earth-like planets around their M-host stars for a fixed resolving power of 100. The error bars show two sigma (standard deviation), single transit errors for selected spectral features, and selected M-dwarf planets at 10 parsecs with optimised wavelength bins. In black: calculated using high resolution JWST NIRSpec BOTS disperser and filter. In grey: calculated using NIRCam weak lens filters and high resolution NIRCam grism filters. Central wavelengths and bandwidths are calculated individually to distinguish between spectral absorption features (see Sec.~\ref{sec:features_10pc}).}
     \label{figure:filter_jwst}
\end{figure}

\subsection{Planetary S/N}
\subsubsection{Signal-to-noise ratio with constant resolution at 10~pc}
For the bottom panel of Fig.~\ref{figure:transmission_snr} we combined all available JWST NIRSpec and MIRI filter-disperser combinations to calculate the highest corresponding S/N of the planetary spectral features during one transit. The specifications of all JWST instruments used in this study can be found in Appendix \ref{sec:appendix}. As already discussed in Sec.~\ref{sec:stellar_snr} certain wavelength regions are not observable with low or medium resolution for bright stars due to saturation effects. Scenarios for GJ~1214, Proxima Centauri, and TRAPPIST-1 are observable with NIRSpec medium resolution spectroscopy at 10~pc below 5~$\mu$m. GJ~644 at 10~pc is only observable at wavelengths longer than 3~$\mu$m. All other M dwarfs can also be observed below 5~$\mu$m via high resolution spectroscopy. For wavelengths longer than 5~$\mu$m MIRI specifications are used to calculate the S/N. The Sun placed at 10~pc is only observable with MIRI MRS, with roughly one-third of the quantum efficiency up to 10~$\mu$m compared to MIRI LRS \citep{rieke2015b,glasse2015}.\\
An increasing stellar radius increases the S/N of the star (Fig.~\ref{figure:comp2pan}) but decreases the transit depth of the planet (Table~\ref{table:planets}). The effective height of the CO$_2$ feature at 4.3~$\mu$m differs only by 10~km between early and late M-dwarf planets owing to similar CO$_2$ abundances of the simulated atmospheres (see top panel of Fig.~\ref{figure:transmission_snr}). The S/N of this feature is however twice as large for late-type M-dwarf planets than for early to mid M-dwarf planets due to the smaller stellar radius. Despite the high stellar signal, the GJ~644 planet has the lowest S/N owing to the exceptionally large radius of the host star. The consideration of the readout noise and the reduction of total exposure time due to the increased number of detector readouts (shorter duty cycle) has an impact on the S/N of bright targets. For GJ~644 at 10~pc half of the transit time is needed for detector readouts using the filter-disperser G395H/F290LP. For fainter targets like TRAPPIST-1 the duty cycle increases. Only 3\% of the transit time is lost for TRAPPIST-1 at 10~pc due to detector readouts. Hence, the detectability of the spectral features is further improved for close mid to late-type M-dwarf planets. For late-type M-dwarf planets we expect the highest S/N compared to earlier M-dwarf planets \citep[see e.g.][]{dewit2013}. The hypothetical Earth-like planet around the M5.5 star Proxima Centauri, placed at 10~pc, has a higher S/N than the planet around the M8 star TRAPPIST-1 placed at 10~pc for all spectral features at wavelength shorter than 3~$\mu$m. Comparing the theoretical radii from \citet{reid2005}, TRAPPIST-1 has a slightly bigger radius and Proxima Centauri a smaller radius than expected for this spectral type. The spectra of very low mass stars show strong atomic lines mainly from \ion{K}{i} and \ion{Ti}{i} between 1~$\mu$m and 1.3~$\mu$m, water vapour bands in the near-IR and CO bands at 2.3 ~$\mu$m to 2.4~$\mu$m \citep{allard2001}. These bands coincide with the absorption features found for the modelled Earth-like atmospheres and decrease the S/N at these wavelengths. We note that M dwarfs later than M6 are affected by dust formation, which increases the uncertainty of the spectral models \citep{allard2012}. \\
As mentioned in Sec.~\ref{sec:transmission_spectra} we do not consider the effect of clouds and hazes on the transmission spectra. Since we use the geometric transit depth (i.e. the planetary radius without atmosphere) as reference to calculate the S/N, the increase of the effective height from additional absorption, scattering, or reflection would not influence the S/N of the spectral bands of, for example CH$_4$ or H$_2$O, at low wavelengths.\\
The atmospheres of Earth-like planets around mid to late M dwarfs have much higher S/N at most absorption features compared to early M-dwarf planets. Nevertheless, at 10~pc a single transit is not enough to detect any spectral feature using JWST with a fixed resolution of 100 due to the maximum S/N of only around 1.5. In the next section we want to investigate whether absorption bands would be detectable when taking into account the entire spectral feature.

\subsubsection{S/N for spectral features at 10~pc}
\label{sec:features_10pc}
 In Fig. \ref{figure:filter_jwst} we show transmission spectra with improved detectability of each spectral features by calculating the bandwidth and central wavelength of each spectral feature. We maximise the lower limit of the error bar (lower value at the two sigma deviation) by shifting the central wavelength and increasing the bandwidth until the optimal value is found. Increasing the bandwidth decreases the noise contamination. But if the selected band spans over a wider wavelength range than the absorption feature, the mean effective height decreases. 
 We note that for a nonsymmetric-Gaussian distribution of the absorption feature, the central wavelength is not necessarily the wavelength of the peak absorption. The selected bandwidth and central wavelength also depend on the overall S/N. For low S/N it is favourable to integrate over a broad wavelength range, including the wings of the spectral feature to decrease the noise contamination. If the noise is low, the bandwidth can be narrow to maximise the transit depth of the absorption feature. To detect a spectral feature it needs to be separated from an atmospheric window with a low  effective height. We find the bandwidth and central wavelength of atmospheric windows by minimising the upper limit of the error bar (upper value at the two sigma deviation) between two absorption features. \\
 \begin{figure}
\centering
   \includegraphics[width=6.5cm]{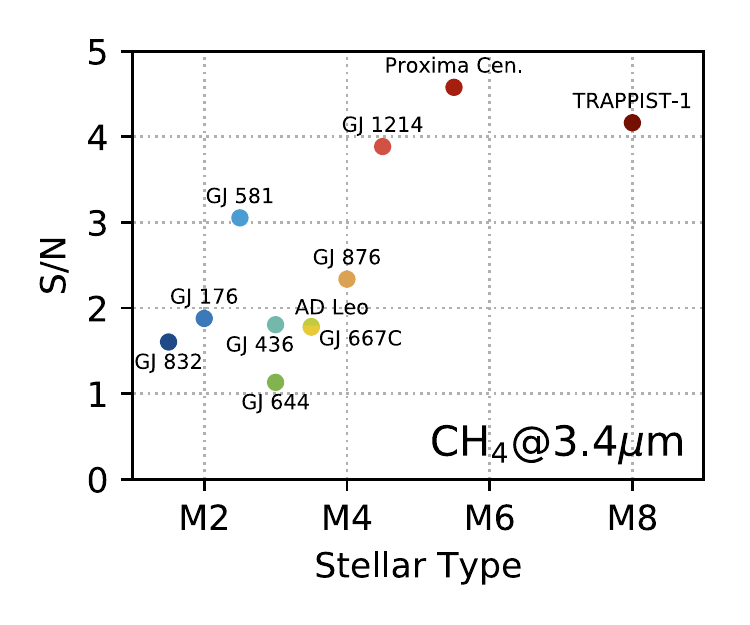}
     \caption{Signal-to-noise ratio of the CH$_4$ feauture at 3.4~$\mu$m for a single transit of each M-dwarf planet at 10~pc.}
     \label{figure:snr_mtype}
\end{figure}
\begin{table}
\centering
\caption{Mean central wavelengths, $\lambda_c$, bandwidths, $\Delta\lambda$ and resolving power for selected spectral features and an S/N of 5. The NIRSpec HRS and MIRI LRS are used to calculate the individual wavelengths and bandwidth for each M-dwarf planet and feature. Values are nearly independent of the distance to the star but dependent on the atmosphere of the planet and the throughput of the filter-disperser combination.}              
\label{table:wavlc_delta}      
\centering                                     
\begin{tabular}{l r r r}          
\hline\hline                        
Spectral feature        & $\lambda_c$[$\mu$m]   & $\Delta\lambda$[$\mu$m]       & $\lambda_c$/$\Delta\lambda$  \\
\hline 
H$_2$O at 1.4 $\mu$m    &       1.38            &       0.06                    & 23.0 \\
CH$_4$ at 2.3 $\mu$m    &       2.33            &       0.15                    & 15.5 \\
CH$_4$ at 3.4 $\mu$m    &       3.37            &       0.32                    & 10.5 \\
CO$_2$ at 4.3 $\mu$m    &       4.27            &       0.11                    & 38.8 \\
H$_2$O at 6.3 $\mu$m    &       6.12            &       0.90                    & 6.8 \\
O$_3$ at 9.6 $\mu$m     &       9.64            &       0.44                    & 21.9 \\
\hline 
\end{tabular}
\end{table}
 Figure~\ref{figure:filter_jwst} shows the detectability of the spectral features using NIRSpec in black and NIRCam in grey. For NIRCam we use five fixed filters in weak lens mode between 1.3~$\mu$m and 2.2~$\mu$m. We omit the narrow-band filters owing to the very low resulting S/N. At longer wavelengths we use the NIRCam grism mode to calculate the error bars. The advantage of using NIRCam would be that observations of one weak lens filter at short wavelengths could be performed simultaneously with one filter-grism at longer wavelengths.\\
 No feature would be distinguishable from an atmospheric window with a single transit for early M-dwarf planets at 10~pc except for the planet around GJ~581, which has a small radius. For mid to late M dwarfs several absorption features could be distinguishable from an atmospheric window within two sigmas with just a single transit. These are CH$_4$/H$_2$O at 1.15~$\mu$m and 1.4~$\mu$m; the CH$_4$ at 1.8~$\mu$m, 2.3~$\mu$m, and 3.4~$\mu$m; and the H$_2$O/CO$_2$ at 2.7~$\mu$m. The CO$_2$ feature at 4.3~$\mu$m could be separated from one of the atmospheric windows below 1.7~$\mu$m if simultaneous observations of both wavelength ranges were to be provided. \citet{barstow2016} showed that just a few transits would be necessary to detect CO$_2$ at 4.3~$\mu$m for an Earth around an M5 star at 10~pc. For an Earth the detection of the H$_2$O and CO$_2$ feature at 2.7~$\mu$m would require more than ten transits due to the lower abundances of H$_2$O in the Earth's middle atmosphere compared to our calculated amounts.\\
Figure~\ref{figure:snr_mtype} shows that the planets around the later M dwarfs with a distinguishable CH$_4$ feature at 3.4~$\mu$m reach an S/N of at least 3. For real observations we expect additional sources of variability like stellar spots \citep{pont2008,zellem2017,rackham2018} and exozodiacal dust \citep{roberge2012,ertel2018}. For this study we consider an S/N of 5 to be sufficient to detect a spectral feature. 

\subsubsection{Detectability of spectral features up to 100~pc}
\begin{figure*}
\centering
   \includegraphics[width=18.5cm]{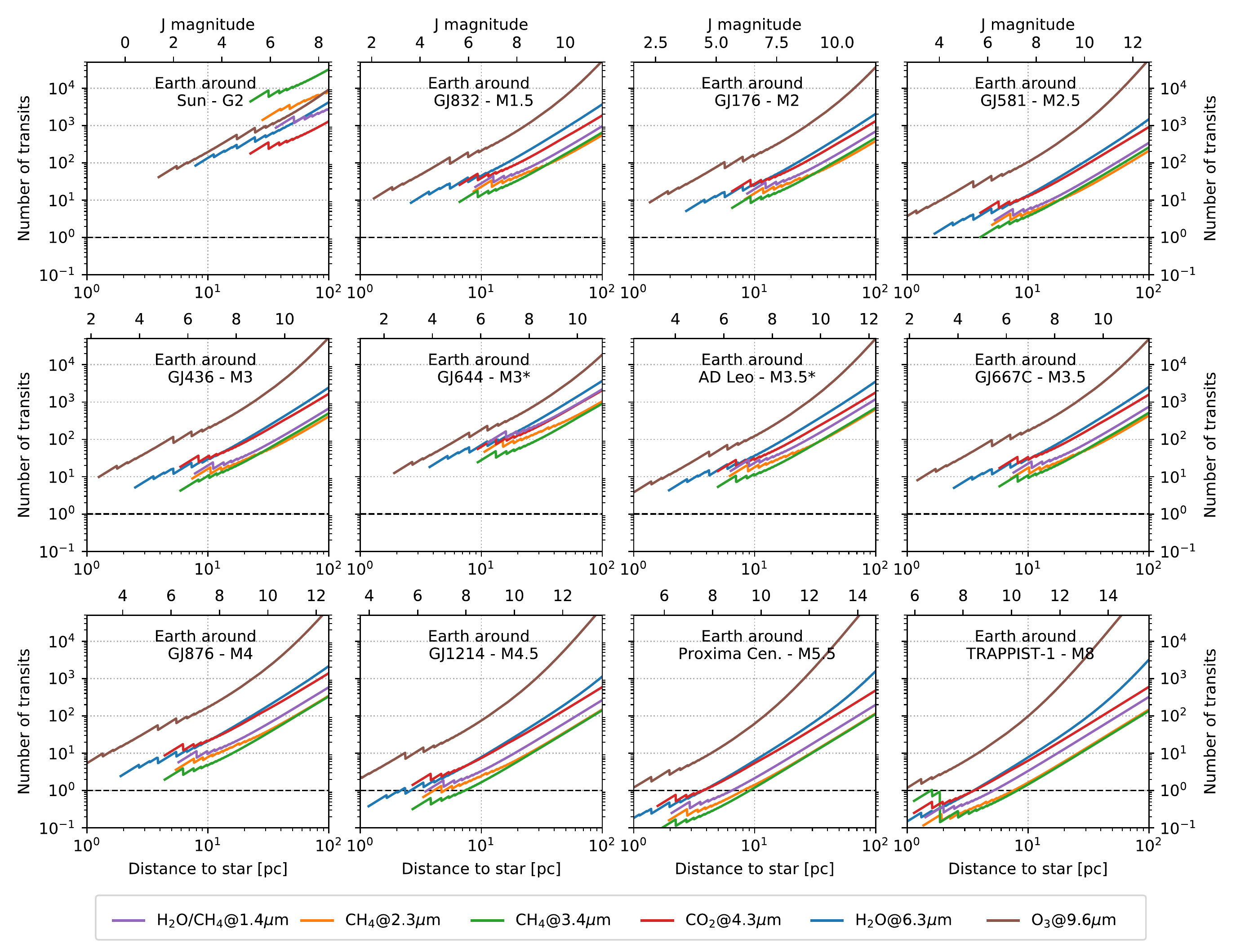}
     \caption{Number of transits needed to reach an S/N of 5 for selected spectral features of M-dwarf planets and the Earth-Sun case and different distances to the star. Central wavelength and bandwidths are shown in Table~\ref{table:wavlc_delta}. The sudden reductions of the needed observation time happen when there are changes in the duty cycle and when a filter-disperser combination with a higher throughput can be used.}
     \label{figure:tint_dis2star}
\end{figure*}
\begin{figure*}
\centering
   \includegraphics[width=15cm]{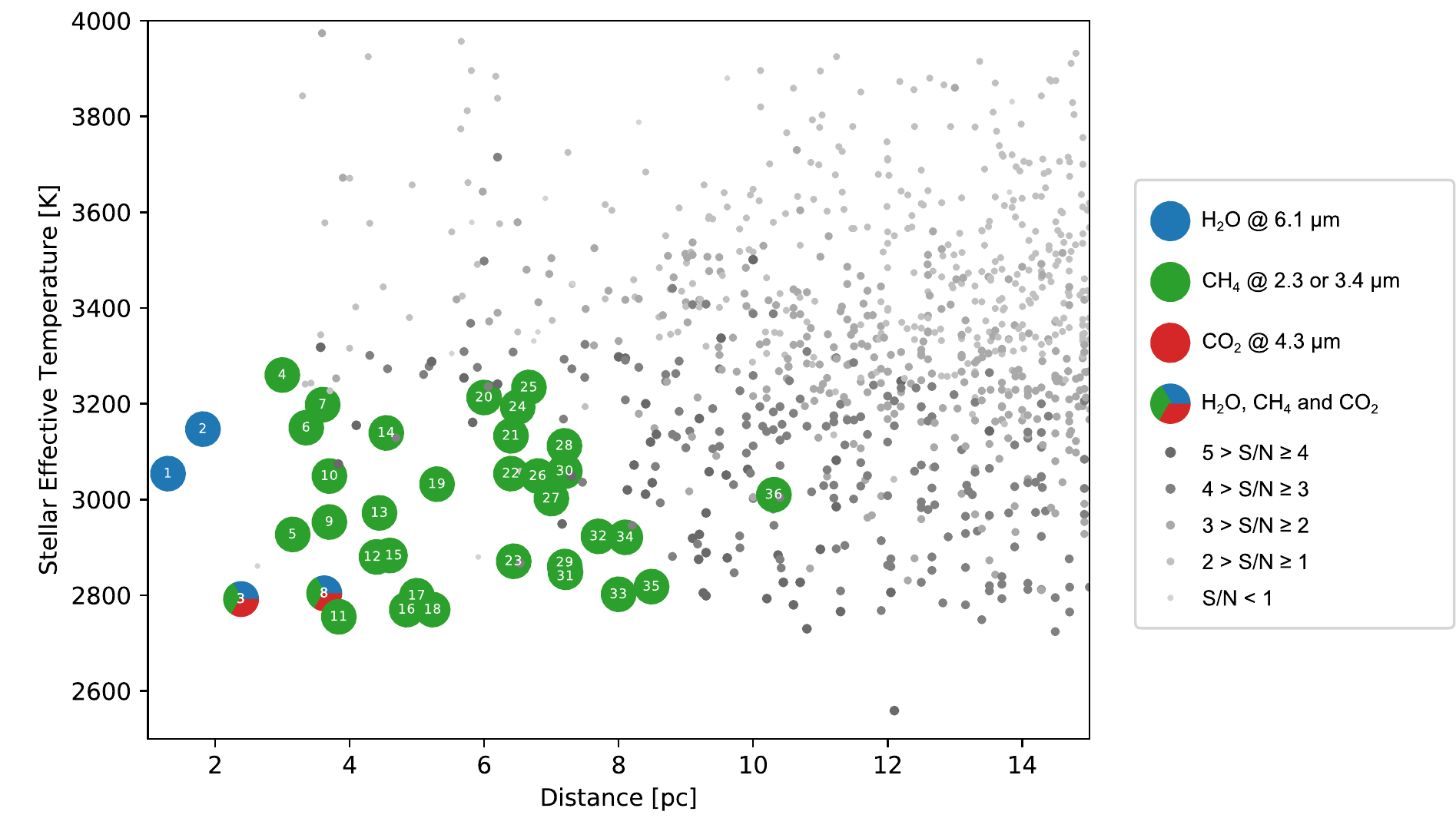}
     \caption{Detectability of hypothetical nearby Earth-like planets around M dwarfs with a single transit. Coloured dots show planets with an S/N of at least 5 for the corresponding spectral features (H$_2$O: blue, CH$_4$: green and CO$_2$: red). More than one colour means that multiple spectral features can be detected with a single transit. The sizes of the grey dots indicate the highest S/N of all spectral features. The numbers inside the coloured dots correspond to the following host stars: 1:~Proxima~Centauri, 2:~Barnard's~star, 3:~Wolf~359, 4:~Ross~154, 5:~Ross~248, 6:~Ross~128, 7:~GJ~15~B, 8:~DX~Cancri, 9:~GJ~1061, 10:~YZ~Ceti, 11:~Teegarden's~star, 12:~Wolf~424, 13:~L~1159-16, 14:~GJ~1245~B, 15:~GJ~1245~A, 16:~GJ~412~B, 17:~LHS~2090, 18:~GJ~1116, 19:~LHS~1723, 20:~GJ~1005, 21:~L~43-72, 22:~GJ~3737, 23:~GL~Virginis, 24:~GJ~1128, 25:~LSPM~J2146+3813, 26:~Ross~619, 27:~G~161-7, 28:~GJ~4053, 29:~G~141-36, 30:~SCR~J0740-4257, 31:~GJ~1286, 32:~LHS~1070, 33:~SCR~J0838-5855, 34:~NLTT~40406, 35:~GJ~3146, 36:~LSPM~J0539+4038. }
     \label{figure:teff_dis2star}
\end{figure*}
The optimal central wavelengths and bandwidths to reach an S/N of 5 are independent from the distance to the star and differ just by a few nanometres between the M-dwarf planets. Table~\ref{table:wavlc_delta} shows the mean central wavelengths, bandwidths, and resolving power for each spectral feature. \\
In Fig.~\ref{figure:tint_dis2star} we show the number of transits necessary to reach an S/N of 5 for the selected spectral features at the given central wavelengths and resolving power. The saturation limit is reached if the minimum integration time is equal to the frame time. We do not consider partial saturation as proposed by \citet{batalha2018}. The saturation limit to observe the entire wavelength range is about 6.3 at J band. Early M dwarfs reach this limit at around 6-10~pc. Mid- to late-type M dwarfs are fully observable at 1-3~pc. A G2 star like the Sun is observable at short wavelengths only beyond a distance of about 30~pc. The O$_3$ feature at 9.6~$\mu$m and the H$_2$O feature at 6.3~$\mu$m can be observed at shorter distances using MIRI MRS. We note that the sudden reductions of the needed observation time are due to changes in the duty cycle (more frames per integration) as shown by \citet{nielsen2016} and changes of the considered filter-disperser combination. We always plot the lowest number of transits needed for detection in cases in which a feature can be observed with several filter-disperser combinations.\\
Because of the high saturation limit and lower S/N (Fig.~\ref{figure:snr_mtype}), none of the spectral features of Earth-like planets around early M dwarfs would be observable with a single transit. For Earth-like planets around late-type M dwarfs lying closer than 4~pc, all selected spectral features except O$_3$ are detectable with a single transit.
By co-adding 10 transits, CH$_4$ is detectable between the saturation limit and 10~pc for most M-dwarf planets. \citet{rauer2011} found that at a distance of 4~pc the CH$_4$ absorption band of an Earth-like atmosphere around an M5 star would reach an S/N of 3 using JWST observations. Owing to the consideration of a broader bandwidth and by taking into account realistic throughputs, our new analysis shows an improved detectability of CH$_4$. For the planet around the M4.5 star, GJ~1214, with a comparable atmosphere to the Earth-like planet around the M5 star modelled by \citet{rauer2011}, we find that even at a larger distance of 7~pc, a single transit is sufficient to reach an S/N of 5. \\
The O$_3$ feature at 9.6~$\mu$m is more difficult to detect because of the low stellar signal at longer wavelengths and the low abundances of O$_3$ of Earth-like planets around inactive M dwarfs. The planet around the active M-dwarf AD~Leo has a similar O$_3$ column as the present Earth. If AD~Leo were to have a transiting Earth-like planet in the HZ, around 50 transits or more than three years of total observing period would be required to detect O$_3$ with JWST. This is in agreement with the results of \citet{rauer2011}. Only 10 transits are necessary to detect O$_3$ at 4~pc for the hypothetical Earth-like planets around the later type M-dwarfs Proxima Centauri and TRAPPIST-1, which have a lower O$_3$ concentration than the planet around AD~Leo.\\
All modelled atmospheres of planets around different M dwarfs show an increased amount of H$_2$O in the middle atmosphere compared to the Earth-Sun case (Fig. \ref{figure:chemical_profiles}). The H$_2$O feature at 6.3~$\mu$m is detectable for mid to late M-dwarf planets at 10~pc by co-adding 10 transits. We find that 20 transits are necessary to detect H$_2$O in the atmosphere of the hypothetical Earth-like planet around the M2.5 star, GJ~581, at 12~pc (J-band magnitude~8). This is comparable to results from \citet{kopparapu2017}, who showed that a detection of a 10 ppm H$_2$O feature observing an M3 star would require around 30 transits at J-band magnitude~8. \\
Previous studies investigated the potential detectability of the TRAPPIST-1 planets using predefined atmospheres \citep[see e.g.][]{barstow_irwin2016,morley2017,kopparapu2017,krissansen2018}.
We calculate the detectability of an Earth-like planet around TRAPPIST-1 considering the effect of the stellar spectrum on the atmosphere of the planet. At least 9 transits are required to detect CO$_2$ of the Earth-like planet around TRAPPIST-1 at 12.1~pc. \citet{morley2017} found that a 1~bar Venus-like atmosphere of TRAPPIST-1e (TRAPPIST-1f) can be characterised by co-adding 4 (17) transits. The O$_3$ feature at longer wavelengths requires the observation of more transits for a detection. \citet{barstow_irwin2016} concluded that 30 transits are needed to detect O$_3$ at Earth abundances for TRAPPIST-1c and TRAPPIST-1d. Our model calculation shows that the O$_3$ levels would be reduced for an Earth-like planet around TRAPPIST-1. Furthermore, at longer wavelengths there is a significant contribution of zodiacal noise for faint targets. Hence, we find that 170 transits are needed to detect O$_3$, which is much more than found by \citet{barstow_irwin2016}, who did not consider zodiacal noise. The time span over which the Earth-like planet in the HZ around TRAPPIST-1 needs to be observed for 170 transits would be more than two years, assuming white noise only and that each transit would improve the S/N perfectly. Hence, it would be difficult to detect O$_3$ with JWST in the atmosphere around an Earth-like planet by transmission spectroscopy, even for a close late-type M-dwarf like TRAPPIST-1. The detectability of CH$_4$, H$_2$O, and  CO$_2$ is improved for M-dwarf planets compared to the Earth because of the larger concentrations of these species in their atmospheres. Up to a distance of about 10~pc CH$_4$, H$_2$O, and CO$_2$ would be detectable in the atmosphere of Earth-like planets orbiting mid- to late-type M-dwarf planets with 10 transits. In Sec.~\ref{sec:tess_mdwarfs} we show for which potential TESS findings within 15~pc it would be possible to characterise the Earth-like planetary atmospheres. 

\subsubsection{Detectability of spectral features of hypothetical TESS findings}
\label{sec:tess_mdwarfs}
The TRAPPIST-1 planets will be among the main future targets for atmospheric characterisation. The TESS satellite will find more transiting exoplanets in the solar neighbourhood \citep{barclay2018}. We used the TESS catalogue of cool dwarf targets \citep{muirhead2018} to calculate the S/N of potential TESS findings. From this catalogue we selected all M dwarfs within 15~pc. The distances themselves are not included in the catalogue. Hence, we took the distance of the M dwarfs in the northern sky from \citet{dittmann2016} and in the southern sky from \citet{winters2014}. We complemented the distances with the data from the CARMENES input catalogue \citep{cortes2017} and with data from \citet{lepine2011}. For values which are present in two or more catalogues, we used the most recent reference. For consistency we used the values for the M dwarfs shown in Table~\ref{table:stars}. The spectra until 5.5~$\mu$m were taken from the PHOENIX model\footnote{phoenix.astro.physik.uni-goettingen.de/} \citep{husser2013}. At wavelengths longer than 5.5~$\mu$m we extended the spectra with NextGen \citep{hauschildt1999}. Since NextGen and PHOENIX have a fixed grid for effective temperature, log~$g$ and [Fe/H], we took the spectra assuming a log~$g$ of 5~cgs and a [Fe/H] of 0.0~dex and linear interpolate the temperature grid to the corresponding value.\\
Figure~\ref{figure:teff_dis2star} shows which hypothetical Earth-like planets around M dwarfs within 15~pc would have a detectable feature for a single transit observation using JWST. To calculate the S/N of each spectral feature we used the modelled atmosphere of the planets around the M dwarfs with the most similar effective temperature (see Table \ref{table:stars}). For 36 of the 915 hypothetical M-dwarf planets at least one spectral feature would however be detectable with a single transit observation. The saturation of the NIRSpec HRS limits the detectability of the planetary atmosphere to stellar effective temperatures of lower than about 3300~K and distances larger than about 8~pc. Bright targets can only be observed at longer wavelengths with MIRI. Proxima~Centauri and Barnard's~star, which are the two closest M dwarfs,  are too bright for NIRSpec HRS. The H$_2$O feature at 6.3~$\mu$m of a hypothetical transiting Earth-like planet would be detectable using MIRI. For the planets around late-type M dwarfs Wolf~359 and DX~Cancri, CO$_2$, H$_2$O, and CH$_4$ would be detectable with a single transit. Because of the large CH$_4$ abundances, the detection of CH$_4$ in the atmosphere of a hypothetical Earth-like planet around mid- to late-type M dwarfs would be possible up to about 10~pc. Three of the 36 M dwarfs have confirmed planets found by radial velocity measurements (Proxima Centauri, Ross~128, and YZ~Ceti). Grey dots show the S/N of the strongest spectral feature. We find 267 hypothetical planets with an S/N of at least 3 for one or more spectral features, including two host stars with confirmed transiting planets (TRAPPIST-1 and GJ~1132) and two early M dwarfs, GJ~581 and Ross~605. For these planets, at least one spectral feature could be detected by co-adding 5 transits. \\
Table~\ref{table:ntr_transit_planets} shows the number of transits required to detect H$_2$O, CH$_4$, CO$_2$, and O$_3$ assuming an Earth-like planet around the 11 closest M dwarfs with a confirmed transiting planet\footnote{exoplanet.eu}. O$_3$ would not be detectable with JWST for any of the synthetic planets owing to the long observation time required. The detection of CH$_4$ would be possible with JWST in a reasonable amount of observation time for an Earth-like planet in the HZ around GJ~1132, TRAPPIST-1, GJ~1214, and LHS~1140.
\begin{table}
\centering
\caption{Number of transits, $N_\text{Tr}$, required to detect H$_2$O, CH$_4$, CO$_2$, and O$_3$ with JWST for an Earth-like planet in the HZ of the corresponding host star. All values have been rounded up to the next integer.}              
\label{table:ntr_transit_planets}      
\centering                                     
\begin{tabular}{l r r r r r}          
\hline\hline                
Host star       &       H$_2$O  &       CH$_4$  &        CO$_2$ &       O$_3$   \\
\hline    
GJ 436          &       31      &       10      &       30      &       209     \\
GJ 1132         &       14      &       4       &       13      &       142     \\
TRAPPIST-1      &       12      &       3       &       10      &       172     \\
GJ 1214         &       17      &       4       &       15      &       171     \\
LHS 1140        &       25      &       5       &       21      &       354     \\
GJ 3470         &       91      &       17      &       80      &       792     \\
NLTT 41135      &       47      &       9       &       41      &       711     \\
K2-18           &       93      &       18      &       80      &       824     \\
LHS 6343        &       167     &       33      &       140     &       1332    \\
Kepler-42       &       168     &       30      &       106     &       2880    \\
K2-25           &       233     &       42      &       184     &       5887    \\
\hline 
\end{tabular}
\end{table}

\section{Summary and conclusions}
We use a range of stellar spectra to investigate the influence of the SED on the atmosphere of Earth-like planets in the HZ around M dwarfs and their detectability by JWST via transmission spectroscopy. The simulations of M-dwarf planets show an increase of CH$_4$, H$_2$O, and CO$_2$ in the middle atmosphere compared to the Earth around the Sun. The abundances of O$_3$ increase for planets around active M dwarfs and decrease for planets around inactive M dwarfs. The results of the atmosphere simulations are in agreement with previous model studies \citep{segura2005,rauer2011,grenfell2013,grenfell2014,rugheimer2015}. For Earth-like planets around early- to mid-type M dwarfs the stellar insolation needs to be downscaled to reproduce the surface temperature of the Earth \citep[see also][]{segura2005}. For late M-dwarf planets like the hypothetical Earth-like planet around TRAPPIST-1 we find that the chemistry feedback requires an upscaling of the stellar insolation to reach the surface temperature of the Earths.\\
One of the main goals of this study was to investigate the detectability of spectral features of the simulated atmospheres. We use a sophisticated S/N model including readout noise contribution and consideration of saturation limits and the duty cycle for each target. Although the model is applicable to any current or future telescope, we limit our analysis to transmission spectroscopy with JWST. We verify the results of our S/N model with the JWST ETC, Pandeia \citep{pontoppidan2016}. We calculate the optimal central wavelengths and bandwidths to detect main spectral features of an Earth-like planet. \\
The detectability of atmospheric features of Earth-like planet around mid to late M dwarfs is much higher than for planets around early M dwarfs. The saturation limits of JWST allow the observation of almost all mid- to late-type M dwarfs at low wavelengths in the solar neighbourhood. For an early-type M dwarf at 10~pc, the integration time due to detector readouts can be reduced up to 50\%. In comparison, for an observation of TRAPPIST-1 the detector readout time is just a few percent of the full observation time when using NIRSpec MRS or HRS. Another advantage of mid to late M-dwarf planets are the increased abundances of biosignature related compounds like CH$_4$ and H$_2$O and the extended atmosphere due to warmer temperatures from the upper troposphere to the mesosphere. This allows a detection of CH$_4$ with only two transits and H$_2$O and CO$_2$ with less than ten transits up to a distance of about 10~pc.  \\
We find that H$_2$O, CO$_2$, and CH$_4$ could be detected in the atmosphere of a hypothetical Earth-like planet around Wolf 359 and DX Cancri with only a single transit. Within 15~pc there are 267 M dwarfs for which detection of at least one spectral feature would be possible by co-adding just a few transits. For hypothetical Earth-like planets around GJ~1132, TRAPPIST-1, GJ~1214, and LHS~1140 a detection of CH$_4$ would require less than ten transits. The predictions of \citet{barclay2018} have shown that terrestrial planets around mid to late M dwarfs within 15~pc are expected to be found with TESS. In this study we conclude that atmospheric features of these planets could be partly characterised with JWST.


  
\begin{acknowledgements}
This research was supported by DFG projects RA-714/7-1, GO 2610/1-1 and SCHR 1125/3-1. We acknowledge the support of the DFG priority programme SPP 1992
"Exploring the Diversity of Extrasolar Planets (GO 2610/2-1)". We thank the anonymous referee for the helpful and constructive comments.
\end{acknowledgements}

\begin{appendix}
\section{JWST configurations}
\label{sec:appendix}
Currently operating telescopes are likely unable to detect spectral features of an Earth-like planet within a reasonable amount of observation time. This paper aims to investigate whether JWST will be able to detect biosignatures and related compounds within a few transits using transmission spectroscopy. The JWST will provide a telescope aperture of 6.5~m diameter and a collecting area of 25.4~m$^2$ \citep{contos2006}. 
For our science case we chose the observing modes of NIRCam, NIRSpec, and MIRI, which are applicable to a single bright object. 

\subsection{NIRSpec}
\label{sec:appendix_nirspec}
The Bright Object Time Series (BOTS) mode is designed for time-resolved exoplanet transit spectroscopy of bright targets \citep{ferruit2012,ferruit2014}. The NIRSpec BOTS uses a fixed slit (FS) mode with the S1600A1 aperture (1.6~$\times$~1.6 arcsec slit) \citep{beichman2014,birkmann2016a}. For the NIRSpec modes there are three sets of disperser and filter combinations available. For low resolution spectroscopy the PRISM with a resolving power of $\sim$100 covers the full NIRSpec range of 0.6~-~5.3~$\mu$m. Medium and high resolution spectroscopy provides four different disperser-filter combinations with a mean resolving power of $\sim$1,000 and $\sim$2,700 respectively (medium resolution: G140M/F070LP, G140M/F100LP, G235M/F170LP, G395M/F290LP; high resolution: G140H/F070LP, G140H/F100LP, G235H/F170LP, G395H/F290LP). For high resolution observations a physical gap between the detectors results in gaps in the wavelength coverage. A subarray of 512~$\times$~32 pixel (SUB512) is used for the PRISM. For medium and high resolutions a subarray of 2024~$\times$~32 pixel (SUB2048) is used in order to cover the full spectrum. \\
To increase the brightness limits it is possible to observe only half of the spectrum (SUB1024A and SUB1024B). Using SUB2048 M dwarfs have a J-band brightness limit of 6.4~$\pm$~0.5 for the G140H and G235H disperser and a limit of ~5.8~$\pm$0.5 for the G395H disperser \citep{nielsen2016}. In this study we also consider the 1024~$\times$~32 pixel subarrays, when the brightness limit for SUB2048 is reached. \\
Each subarray has a different single frame time, which is the same as the readout time. The frame time for SUB512 is 0.226~s, for SUB1024A and SUB1024B 0.451~s and for SUB2024 0.902~s. It is recommended to use as many frames as possible per integration because the S/N does not simply scale with the number of frames owing to changes in the detector duty cycle \citep{rauscher2007,nielsen2016}. The duty cycle increases with increasing number of frames. We use the frame time as a minimum exposure time and do not consider a partial saturation strategy as proposed by \citet{batalha2018}. If one pixel is saturated in a shorter time than the single frame time, the target is considered to be not observable for this filter-disperser combination. \\
For detailed information about NIRSpec we refer to \citet{birkmann2016b} and the on-line documentation\footnote{jwst.stsci.edu/instrumentation/nirspec}. The wavelength ranges, resolving powers, and throughputs of each filter-disperser combination can be found in the on-line documentation, in \citet{birkmann2016a} and in \citet{beichman2014}. Other relevant information such as dark current (0.0092~e$^-$s/pixel), pixel scale (0.1~arcsec/pixel), readout noise (6.6 e$^-$), and full well capacity (55,100~e$^-$) are taken from the on-line documentation. The dispersion curves and PSFs are described in \citet{perrin2012} and were downloaded from the documentation page. 

\subsection{NIRCam}
The NIRCam grism time-series observations allows for simultaneous observation of short wavelengths (1.3~$\mu$m to~2.3~$\mu$m) and longer wavelengths (2.4~$\mu$m to~5.0 $\mu$m) \citep{beichman2014}. To observe bright targets at short wavelengths, a weak lens is used to defocus incoming light. The expected saturation limit change for the strongest wave of defocus 8 (WLP8) in a subarray of 160~$\times$~160 pixel is 6.5 magnitudes \citep{greene2010}. For the medium filter centred at 1.4~$\mu$m this means a saturation limit of $\sim$~4 at K-band for a G2V star, assuming 80\% of the full well capacity (105,750~e$^-$). The readout time and the minimum exposure time is 0.27864~s. We use all medium (R~$\approx$~10) and wide (R~$\approx$~4) filters available for the weak lens mode for our S/N calculations (F140M, F150W, F182M, F210M, F200W). Further information about the NIRCam weak lens mode such as dark current (0.0019~e$^-$s/pixel), pixel scale (0.031 arcsec/pixel), readout noise (16.2~e$^-$), and full well capacity (105,750~e$^-$) are given in \citet{greene2010} and the NIRCam documentation\footnote{jwst.stsci.edu/instrumentation/nircam}. \\
At longer wavelengths a grism is used for slitless spectroscopy between 2.5 $\mu$m~and ~5.0 $\mu$m. The wide filters (F277W, F322W2, F356W, F444W) have a mean resolving power of $\sim$~1,600 and a dispersion of 1~nm/pixel \citep{greene2016,greene2017}. We use the smallest grism subarray of 64~$\times$~2048 pixel and a frame and readout time of 0.34061~s. For an M2V star the brightness limit at K band is $\sim$~4. Information on the wavelength dependent filter throughput and resolving power can be found in \citet{greene2017}. The PSFs were downloaded from\footnote{stsci.edu/$\sim$mperrin/software/psf\_library/} \citet{perrin2012}. Other specifications like dark current (0.027~e$^-$s/pixel), pixel scale (0.063~arcsec/pixel), readout noise (13.5~e$^-$), and full well capacity (83,300~e$^-$) are taken from the on-line documentation.

\subsection{MIRI}
The JWST Mid-Infrared Instrument (MIRI) provides medium resolution spectroscopy (MRS) from 4.9~$\mu$m to  29.8~$\mu$m and low resolution slitted and slitless spectroscopy (LRS) from 5~$\mu$m to 12~$\mu$m \citep{wright2015,rieke2015}. The MRS mode has four channels with three sub-bands each. The resolving power for short wavelengths is 3,710 and decreases for long wavelengths to 1,330 \citep{wells2015}. The brightness limits for MIRI MRS are around magnitude 4 at K band for late-type host stars for a 2 frame integration \citep{beichman2014,glasse2015}. For MRS there is no subarray available. The full size is 1024~$\times$~1032 pixels, which results in a frame time and readout time of 2.775~s. \\
The LRS slitless mode provides the possibility to read out a subarray of 416~$\times$~72 pixels. This decreases the frame and readout time to 0.159~s. Hence, despite the much lower resolving power of $\sim$100 compared to MRS \citep{kendrew2015}, the brightness limit is just 2 magnitudes fainter for the K-band \citep[magnitude 6 at K band;][]{beichman2014,glasse2015}. 
For targets brighter than magnitude 6 at K band, we always use LRS owing to the better throughput of around a factor of 3 up to 10~$\mu$m compared to MRS \citep[][\footnote{http://ircamera.as.arizona.edu/MIRI/pces.htm}]{rieke2015b,glasse2015}. We took other specifications like dark current (0.2~e$^-$s/pixel), pixel scale (0.11~arcsec/pixel), readout noise (14~e$^-$), and full well capacity (250,000~e$^-$) from the on-line documentation\footnote{https://jwst.stsci.edu/instrumentation/miri} and from references therein.

\end{appendix}

\bibliographystyle{aa} 
\bibliography{refs} 

\end{document}